\documentclass[usenatbib]{mn2e}
\usepackage{graphicx}
\usepackage{amssymb}
\usepackage{lscape}
\usepackage{ulem}
\usepackage{txfonts}
\usepackage{url}
\usepackage{subfig}

\def\kms{km~s$^{-1}$}

\def\kpc{$h_{70}^{-1}$ kpc}

\def\ha{H$\alpha$}
\def\hb{H$\beta$}

\def\delv{$\Delta v$}
\def\deloh{$\Delta$log(O/H)}
\def\delsfr{$\Delta$log(SFR)}
\newcommand{\nii}{[N{\sc ii}]}
\newcommand{\oii}{[O{\sc ii}]}
\newcommand{\oiii}{[O{\sc iii}]}

\newcommand{\rp}{$r_p$}
\title[Galaxy Pairs in the Sloan Digital Sky Survey - V]{Galaxy Pairs in the Sloan Digital Sky Survey - V. Tracing changes in star formation rate and metallicity out to separations of 80 kpc.}

\author[Scudder et al.] {Jillian M. Scudder$^1$\thanks{jscudder@uvic.ca}, Sara L. Ellison$^1$, Paul Torrey$^2$, David R. Patton$^3$, J. Trevor Mendel$^1$\\
$^1$ Department of Physics and Astronomy, University of Victoria, Victoria, British Columbia, V8P 1A1, Canada.\\
$^2$ Harvard Smithsonian Center for Astrophysics, 60 Garden St., Cambridge, MA 02138, USA\\
$^3$ Department of Physics \& Astronomy, Trent University, 1600 West Bank Drive, Peterborough, Ontario, K9J 7B8, Canada.}

\begin{document}

\maketitle

\begin{abstract}
We present a sample of 1899 galaxies with a close companion taken from the Sloan Digital Sky Survey Data Release 7.  The galaxy pairs are selected to have velocity differences \delv $< 300$ \kms, projected separations (\rp) $<80$ \kpc, mass ratios between 0.1 and 10, and robust measurements of star formation rates and gas-phase metallicities.  We match the galaxies in total stellar mass, redshift, and local density to a set of 10 control galaxies per pair galaxy.  For each pair galaxy we can therefore calculate the statistical change in star formation rate (SFR) and metallicity associated with the interaction process.
Relative to the control sample, we find that galaxies in pairs show typical SFR enhancements that are, on average, 60\% higher than the control sample at \rp\ $<$ 30 \kpc.    It is at these small separations that the strongest enhancements in SFR  (by up to a factor $\sim$10) are measured, although such starbursts are rare, even amongst the closest pairs.  In addition, the pairs demonstrate more modest SFR enhancements of $\sim$30\% out to at least 80 \kpc~(the widest separations in our sample).  This is the first time that enhanced SFRs have been robustly detected out to such large projected separations.  Galaxies in both  major and minor mergers show significant SFR enhancements at all \rp, although the strongest starbursts (with SFR enhancements of a factor of $\sim$ 10) appear to be found only in the major mergers. We also find evidence that SFR enhancements are synchronised in an interacting pair, such that a higher SFR in one galaxy is accompanied by an increased SFR in its companion.  For the first time, we are also able to trace the metallicity changes in galaxy pairs as a function of projected separation.  The metallicity is generally diluted in galaxy pairs by $\sim 0.02$ dex, with an average metallicity decrement of $-$0.03 dex at the smallest separations, a trend that mirrors the SFR enhancements as a function of \rp.
The SFR and metallicity trends with projected separation are interpreted through a comparison with theoretical models.  These simulations indicate that the peak in SFR enhancements at small separations is due to systems near the end of the merger process.  The extended plateau in SFR enhancements out to at least 80 \kpc~is dominated by galaxies that have made a pericentric passage and are now experiencing triggered star formation on their trajectory towards apogalacticon, or on a subsequent close approach. 
\end{abstract}


\section{Introduction}
Galaxies that experience a close encounter with a companion are expected to undergo significant changes.
Both observations and simulations have been used to probe the internal properties of galaxies as they progress through a merger, with simulations providing a framework through which the observational results may be interpreted.  
Currently, theoretical models present a consistent general picture of the evolution of a galaxy in a merger.   Strong tidal interactions may trigger bar instabilities in the central regions of the galaxy in both the stellar and gaseous components.  These bars are misaligned, and torques exerted on the gas by the stars result in the loss of angular momentum in the gas at larger radii.  This gas then falls towards the nucleus of the galaxy, efficiently funnelled by the bar instabilities  \citep{Barnes1996, Mihos1996,Cox2006, diMatteo2007, Montuori2010, Rupke2010a, Torrey2012}.  
Indirect signatures of this picture are visible in the gas-phase metallicities and star formation rates of the interacting pairs.  Gas inflowing from the outer regions is generally of a lower metallicity than the nuclear regions, so inflow from larger radii results both in a diluted nuclear metallicity \citep{Kewley2006a, Ellison2008, Michel-Dansac2008} and in a flattening of the standard metallicity gradient present in spiral galaxies \citep{Rupke2010b, Kewley2010a, Perez2011a}. 
This new central concentration of gas provides the ideal catalyst for a significant starburst.  Higher than average central star formation rates occur in nearly all simulations of galaxy mergers \citep[e.g.,][]{Mihos1994, Mihos1996, diMatteo2007, Montuori2010}, and are also a ubiquitous feature in observational studies \citep[e.g.,][]{Larson1978, Donzelli1997, Barton2000, Lambas2003, Alonso2004, Alonso2006, Woods2007, Ellison2008, Ellison2010, Darg2010, Xu2010}.  

The strength of the induced star formation can vary dramatically from merger to merger.   A significant contributing factor to this variation is predicted to be the mass ratio of the two galaxies \citep{Cox2006}. 
Galaxies in major mergers, i.e., with masses within a factor of 3, have previously been reported to show the strongest star formation rate (SFR) enhancements, on average, with both galaxies showing comparable enhancements \citep{Woods2007, Ellison2008, Xu2010, Lambas2012}.
Additionally, simulations suggest that a number of parameters within the merger exert a significant influence upon the extent to which gas is funnelled to the central regions of a galaxy.  Most notably, orbital parameters can control the strength of the SFR response \citep{diMatteo2007, DOnghia2010}.  The difference between a prograde--prograde and prograde--retrograde merger can alter the triggered star formation by a factor of 2, and the separations of the nuclei at first passage can inhibit SFR at coalescence if the tidal forces are so strong that gas is preferentially ejected into tidal features, rather than collecting in the central regions.  Gas fractions also seem to play a role, with lower gas fraction galaxies showing weaker SFR triggering \citep{diMatteo2007}.  

Observationally, studies of the SFR in galaxy pairs with projected separation (\rp) are largely in agreement with each other.  Galaxies in pairs show enhanced SFRs out to separations of $\sim$ 30 \kpc, with the strongest enhancement at the smallest separations, and a smooth decline to a fiducial value \citep{Lambas2003, Alonso2006, Nikolic2004, Ellison2008, Li2008}.  Some studies have found evidence for SFR enhancements at wider separations \citep[e.g.,][]{ Barton2000, Lin2007, Robaina2009, Wong2011}, although generally this only weakly extends the trend to $\sim$50 \kpc.    Galaxies are generally expected to be most strongly enhanced just after first passage, and again at coalescence, but the timescales for catching a galaxy during a significant starburst are also tied to the strength of the burst \citep{Torrey2012}.  The highest SFR enhancements are generally of the shortest duration, whereas lower-level enhancements can persist for longer periods of time \citep{diMatteo2007, Montuori2010}. The trend of SFR with projected separation is therefore likely to be a combination of merger-induced changes and the timescale on which the observational snapshot is taken.

In practice, our ability to determine the merger phase of a galaxy pair is severely limited, as our only observable indicators are that of the projected separations (\rp) of the two galaxies, or the morphological disturbance of the galaxy.  However, even this latter metric is fraught with problems.  Depending on the orbit of the encounter, galaxies do not always show strong tidal features after an interaction \citep[e.g.,][]{Toomre1972, DOnghia2010}, and the ability to classify galaxies, either visually or using an automated method, is strongly dependent on the surface brightness of the tidal features relative to the limitations of the data.  With large data sets, such as those obtained from large surveys, the imaging is often not very deep, and the samples are large enough that visual classifications quickly become impractical (although see Darg et al. 2010).  As a result, many studies rely entirely upon the projected separations as a metric for determining the merger phase of a galaxy pair, by correlating large \rp~with a large time elapsed since pericentric passage, and small separations with galaxies actively undergoing a close passage.  
  The SFR enhancement trends with \rp~are thus usually interpreted as the signature of a sharp increase in SFR at pericentric passage, with the strength of the enhancement dwindling as the separations between the galaxies increases \citep[e.g.,][]{Barton2000, Barton-Gillespie2003, Nikolic2004,Woods2010}.  
This physical model would indicate that galaxies are able to rapidly funnel gas to the central regions of a galaxy after an interaction, and that gaseous concentration is rapidly converted to stars, with the SFR declining as the galaxies separate and the burst ceases.

To date, there has been no simultaneous statistical study of both the gas phase metallicity and the star formation rates in a sample of galaxy pairs.  Both the SFR and the gas phase metallicity are expected to change significantly throughout the merger as gas flows are induced through the galaxy; studying both of these quantities simultaneously allows us to gain insight into when changes are induced.  In this work, we will tackle this issue by using a sample of pair galaxies from the Sloan Digital Sky Survey Data Release 7 \citep[SDSS DR7;][]{Abazajian2009}.  Our sample will be used to precisely quantify differences between the pair and the control samples over a wide range of projected separations and mass ratios.    A suite of simulations based on the work in \citet{Torrey2012} is developed to allow us to analyse the trends with \rp~for characteristic markers of the merger stages.

To this end, we use the sample of galaxy pairs in the SDSS DR7 compiled by \citet{Patton2011} and the metallicities calculated in \citet{Scudder2012} to construct a clean pairs sample with stringent quality control measures.    In Section \ref{sec:sample} we review our sample selection and define the control sample, along with a description of our metallicity and SFR values.  In Section \ref{sec:offsets}, we describe our methodology for quantifying the changes in the SFR and the metallicity relative to the control sample and further quantify the changes for a subsample of morphologically disturbed galaxies.  In Section \ref{sec:stats} we describe the diagnostic statistics of the sample in more detail, and in Section \ref{sec:sims}, we compare our results to simulations in order to develop an interpretative framework for our trends with projected separation.  We conclude with a comparison to previous work and an analysis of the physical picture our results suggest in Section \ref{sec:discussion}.

Throughout this work, we assume $\Omega_M = 0.3$, $\Omega_{\Lambda}=0.7$
and H$_0$ = 70 \kms\ Mpc$^{-1}$.

\section{Sample Selection}
\label{sec:sample}
In this section we describe the criteria applied to the spectroscopic pool of galaxies in the SDSS DR7 from which we compile our final pairs and control samples.  We wish to select a tightly controlled set of pair galaxies with a high probability of being a clean selection of physically associated and interacting systems.  We require consistently calculated quantities across the pairs and the controls in order to be able to accurately compare between the samples.

In order to qualify for the pairs sample or the control sample, a galaxy must have a stellar mass\footnote{Unless otherwise stated, all mass values in this work are stellar masses.} and sufficiently strong emission lines so that a gas-phase metallicity and star formation rate can be calculated.     
In order to prevent the issues with Sloan photometry detailed in \citet{Simard2011} from affecting the mass estimates, the masses used here have been recalculated from the MPA catalogue\footnote{Available here: \url{http://www.mpa-garching.mpg.de/SDSS/DR7/}}, using their inferred mass to light ratios (M/L).  The relationship between the MPA-derived M/L and ($g-r$) colour is determined for the total MPA catalogue. This relationship is then used to derive masses from the $g-r$ colours derived from the updated photometry presented in \citet{Simard2011}.  With a new M/L value for each galaxy, the luminosities from \citet{Simard2011} are used to calculate a mass without needing to rely on the Sloan magnitudes.   
We use the SFR values presented in the MPA catalogue, calculated according to the \citet{Brinchmann2004} work. For galaxies with sufficiently high emission line signal to noise (S/N), \citet{Brinchmann2004} uses a set of 6 emission line fits to the SDSS spectra to calibrate the SFR; lower S/N galaxies are calibrated with the strength of the \ha~line.
Given the high S/N we require for emission lines in our sample, all SFR values will be from the multi-emission line fits.  The median $1\sigma$ error on any individual SFR value in our sample is $\sim0.09$ dex.
The SFR values taken from Sloan's 3 arcsecond fibre can be corrected to total values using models fit to the photometry of the galaxy outside the fibre, taking into account the fraction of light not contained within the fibre.  A full description of the aperture correction methodology and bias testing is present in the \citet{Brinchmann2004} work.  However, for this work, we use the fibre values only.

We further require that each galaxy have a consistently calculated metallicity, following the criteria of \citet{Scudder2012}.  In order to ensure that all emission lines needed for the metallicity calibration are within the spectral range of the SDSS, we impose a lower redshift limit of $z>0.02$.  \citet{Kewley2008} demonstrate the problems of comparing between metallicity calibrations; a systematically calculated diagnostic is crucial.  The metallicity calibrations of \citet{Scudder2012} use the adaptation of the \citet{KD02} method presented in \citet{Kewley2008}.  
Briefly,  a S/N $>$ 5 is required in H$\alpha$, H$\beta$, \oii$\lambda3727$, \oiii$\lambda{4959}, \lambda5007$, and \nii$\lambda{6584}$.  Additionally, the Balmer ratio (\ha/\hb) S/N is required to be $>$ 5.  Galaxies must also be classified as star forming on the diagnostic diagram of \citet{bpt}, using the \citet{Kauffmann2003} diagnostic line.  Using duplicate spectra, the standard deviation of the metallicity can be calculated, and results in a median error of $0.015$ dex.  We refer the reader to \citet{Scudder2012} for a complete description of the emission line quality control and metallicity calibrations.  

Our galaxy pairs sample is based upon the spectroscopic catalogue of \citet{Patton2011}.  For a full description of the sample selection algorithm, we refer the reader to that work. Briefly, \citet{Patton2011} select galaxies with a spectroscopic close companion within \rp~$<$ 80 \kpc, with a velocity difference \delv~$<$ 10,000 \kms, and within a mass ratio of 10:1.   This results in a pairs sample of 23,397 galaxies in pairs or higher order multiples\footnote{$\sim$8\% of these galaxies have more than one close companion; 1.6\% have more than 2 companions.  Excluding these from our sample does not significantly impact our results.}.  Our master pairs catalogue is the result of a re-running of the algorithm of \citet{Patton2011} with a minor alteration to the culling process meant to account for fibre collisions in Sloan.  Due to the fixed fibre collision constraint of 55'' \citep{Strauss2002}, galaxies with a projected companion nearer than 55'' will be preferentially missed in the spectroscopic sample.  Some galaxies with separations less than the collision limit do have spectroscopic companions, primarily as a result of overlap between plates.  In order to compensate for this incompleteness effect, 67.5\% of galaxies in pairs with separations $>$ 55$^{\prime\prime}$ are randomly culled from our sample, according to the incompleteness calculations of \citet{Patton2008}.  In \citet{Patton2011}, this cull was done without taking into account the culled galaxy's companion; i.e., a galaxy could remain in the sample while its companion was culled.  For the sample used in this work, the culling has been repeated, but once one galaxy in a pair has been excluded from the sample, its companion is also removed. 

Galaxies with \delv~values at the high end of the distribution are unlikely to be physically associated.  To select a sample of systems which are more likely to be physically associated, a \delv~$ <300$ \kms~limit is imposed to minimize the influence of projected pairs \citep{Patton2000}. This cut is in line with \delv~cuts in other works, which usually range between 250 -- 500 \kms \citep[e.g.,][]{Lambas2003, Ellison2008, Patton2011}.   1,899 galaxies in the master pairs catalogue pass all criteria, and comprise our final galaxy pairs sample\footnote{We do not require that both a galaxy and its companion pass all criteria.}.  

\subsection{Matching to Controls}
In order to determine whether an interaction leads to changes in galactic properties, we must also define a control sample.  To this end, we define a sample of non-pair galaxies as any galaxy which meets the mass, SFR and metallicity requirements of the pairs sample, but is not part of the full pairs catalogue (the ``control pool'').

The largest potential sources of bias between a sample of galaxies and its control are differences in the stellar masses, redshifts, and environments of the galaxies in the two samples \citep{Perez2009}.  In order to eliminate these sources of bias from our sample, we wish to match our control sample in all three parameters, such that the normalized distribution of the final control sample matches that of the pair galaxies.  Simply matching in stellar mass and redshift could bias the results, as it has been shown that pairs tend to exist in higher density environments than non-pairs \citep[e.g.,][]{Barton2007, Patton2011}, and there is a well known trend of SFR with density \citep[e.g.,][]{Kauffmann2004, Poggianti2008, Cooper2008}.  Our density metric is calculated as an overdensity by taking the 5th nearest neighbour density and normalizing by the median density in a redshift slice of $\pm0.01$ around the galaxy.

Galaxies in our pairs sample are therefore matched to galaxies in the control pool simultaneously in mass, redshift, and local density, in a similar way to \citet{Ellison2010}.  Briefly, our algorithm finds the galaxy from the control pool which is the simultaneous best match in all three parameters for each pair galaxy.  After every pair galaxy has been matched to a control, a Kolmogorov-Smirnov (KS) test is used on the total distributions of pairs and control.  If the KS test finds that the distributions are consistent with being drawn from the same parent distribution at $>30\%$, then the tentative controls become part of the control sample.  Matching continues without replacement until 10 control galaxies have been matched to each pair galaxy in our sample, or the KS test returns a probability $<30\%$.  For our sample of galaxy pairs, the algorithm ran without failing the KS criterion; the final control sample therefore contains 18,990 galaxies.  

\begin{figure}
\centerline{\rotatebox{0}{\resizebox{8.7cm}{!}
{\includegraphics{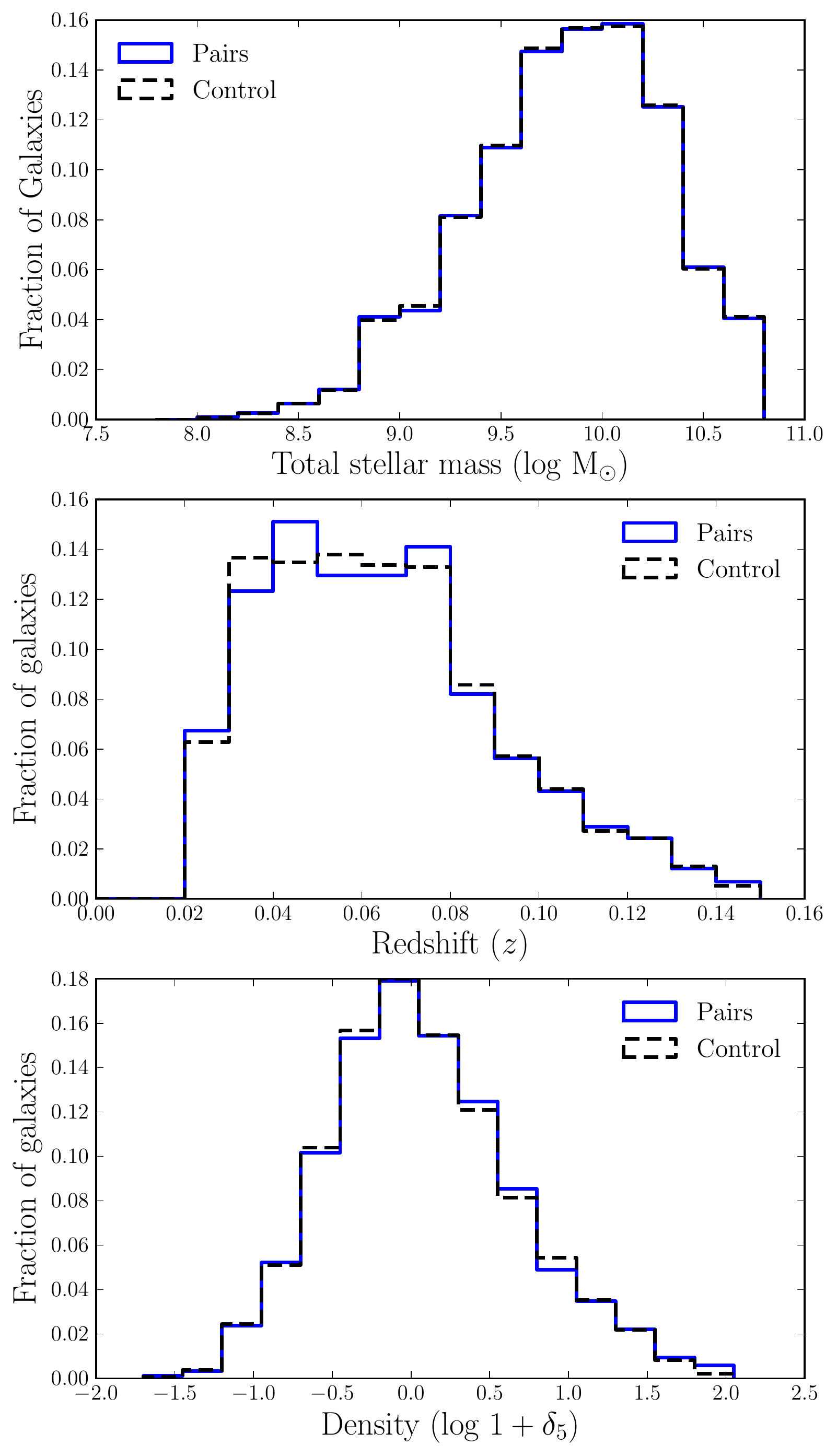}}}}
\caption{\label{fig:sample_dists} Normalized distribution of the pair galaxies (solid blue) and the control sample (dashed black), matched in mass (top), redshift (middle), and density (bottom).  Density is measured via the distance to the 5th nearest neighbour.  Each pair has 10 controls.  KS tests result in a probability of $\sim 99.8\%$ that both the stellar mass and density distributions were drawn from the same parent distribution, and a probability of $\sim 83.8\%$ that the redshift distributions were drawn from the same parent population.}
\end{figure}

The normalized distributions of stellar mass, redshift, and density for the pairs and control samples are displayed in Figure \ref{fig:sample_dists}.  KS tests on the final distributions result in probabilities of 99.77\% for the stellar masses of the two samples,  83.84\% for the redshifts, and 99.85\% for the density distribution, indicating that the galaxy pair and control samples are very unlikely to have been drawn from different parent populations. 
Pairs and their controls are typically matched to within 0.001 dex in total stellar mass, $6\times 10^{-6}$ in redshift, and 0.0008 dex in density, with a median range within the set of control galaxies of 0.07 dex in mass, 0.005 in redshift, and 0.09 dex in density. 

\section{SFR and metallicity offsets}
\label{sec:offsets}

With a well-matched control sample of non-pair galaxies, it is now possible to proceed to compare the SFRs and metallicities of the pairs and controls.  As we are using the fibre values for the SFRs, both our SFRs and metallicities are values for the central 3$^{\prime\prime}$.  Our adopted methodology is the same as \citet{Scudder2012}, which itself is a modification of the method presented in \citet{Patton2011}.
Each galaxy in a pair is matched to 10 controls, which all have metallicity and SFR values.  The median SFR or metallicity of the 10 controls is taken, and subtracted from the value for the matched pair galaxy, as expressed in Equations \ref{eq:mzr_offset} \& \ref{eq:sfr_offset}.  We define the resultant value, \delsfr~or \deloh, to be our `offset' values from the control.
\begin{equation}
\label{eq:mzr_offset}
\Delta \mathrm{log(O/H)} = \left(\mathrm{log(O/H)} + 12\right)_{pair} - \mu_{1/2}\left( \mathrm{log(O/H)} + 12\right)_{controls},
\end{equation}
\begin{equation}
\label{eq:sfr_offset}
\Delta \mathrm{log(SFR)} = \mathrm{log(SFR)}_{pair} - \mu_{1/2}\mathrm{log(SFR)}_{controls},
\end{equation}
where $\mu_{1/2}$ signifies the median. 

Positive offset values indicate enhancements over the control sample, i.e., metallicity enrichment or SFR enhancement, whereas negative offset values indicate suppression relative to the control sample, i.e., metallicity dilution or suppressed SFR.  This method has the advantage of allowing the quantification of the changes in SFR and metallicity on a galaxy by galaxy basis, relative to the 10 control galaxies to which it is matched.  Both SFR and metallicity are known to have strong mass dependences \citep{T04, Ellison2008a, Noeske2007, Elbaz2011}, but as pair galaxies and their controls are tightly matched in mass, a comparison to the matched control galaxies is effectively a comparison at fixed mass.    

\begin{figure}
\centerline{\rotatebox{0}{\resizebox{9cm}{!}
{\includegraphics{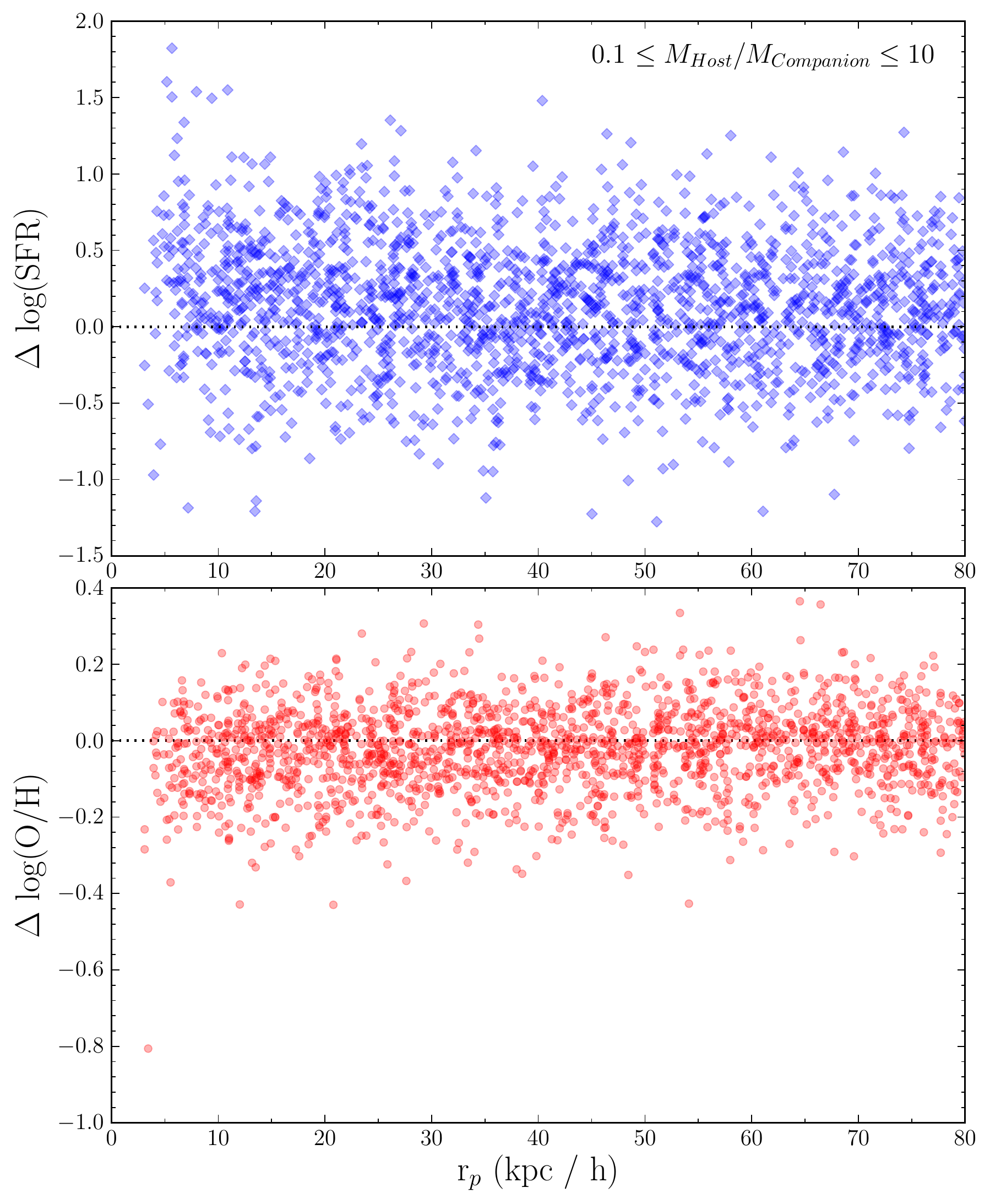}}}}
\caption{\label{fig:all_unbinned} Offset values for all 1899 galaxies in pairs sample as a function of projected separation.  The top panel shows the SFR offsets, and the bottom panel shows metallicity offsets. The horizontal black dotted lines indicate the zero line.}
\end{figure}

\begin{figure}
\centerline{\rotatebox{0}{\resizebox{9cm}{!}
{\includegraphics{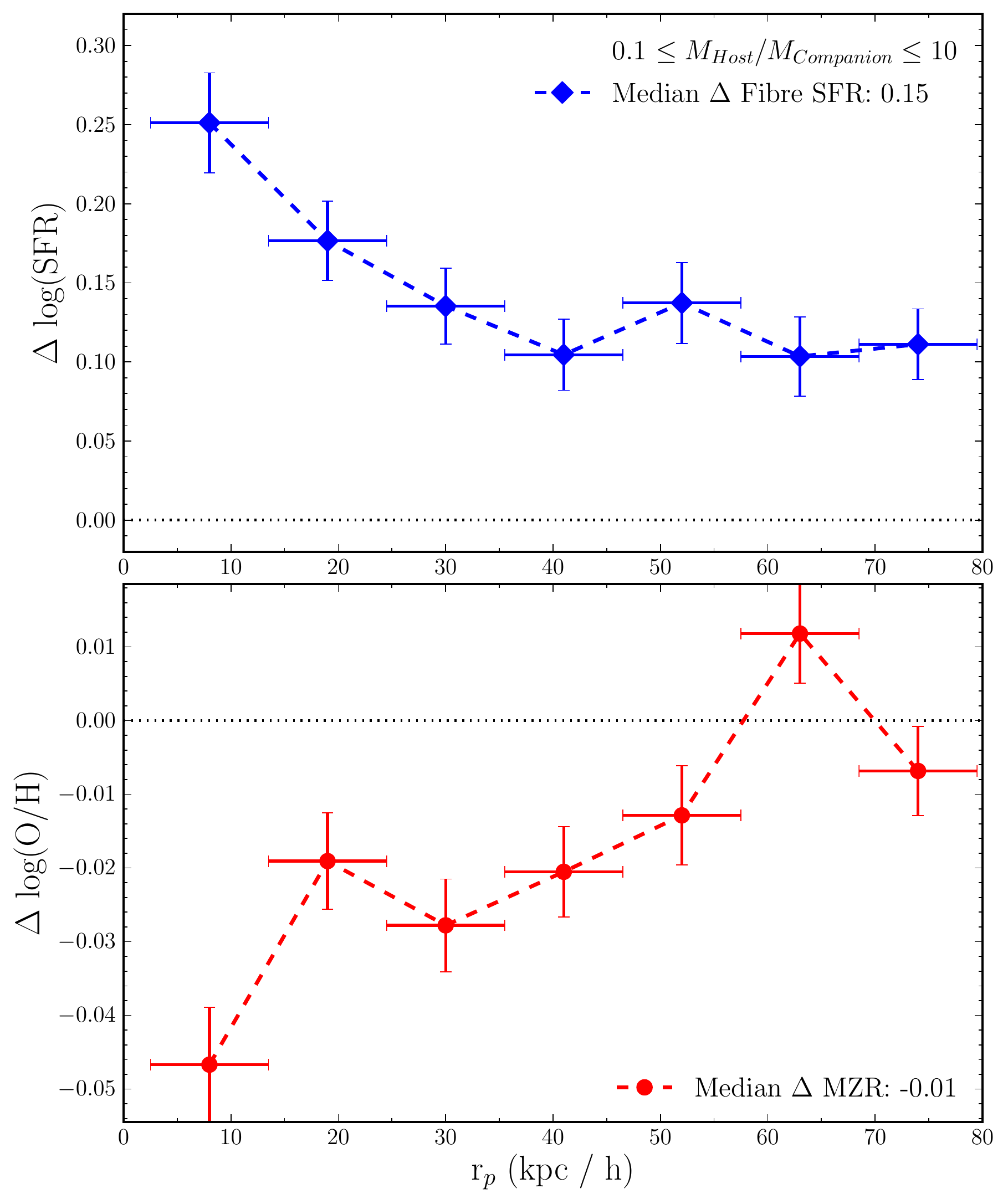}}}}
\caption{\label{fig:all_offset} All 1899 galaxies in pairs sample.  The top panel shows the SFR offsets, bottom panel shows metallicity offsets.  Points are median values for the bin, and error bars are standard error on the median.  Horizontal black dotted lines indicate the zero line.}
\end{figure}

The calculated SFR and metallicity offsets are plotted against projected separation (\rp) in Figure \ref{fig:all_unbinned}.   The distribution of \delsfr~is already visibly offset from zero.  This trend is made clearer in Figure \ref{fig:all_offset}, which shows the same sample of galaxies, but binned in \rp.  SFRs are significantly offset from the control sample out to 80 \kpc, the widest separation probed in our sample, with an increase in the offset magnitude at the smallest separations (\rp $\leq 15$ \kpc) to $\sim 0.25$ dex (about a factor of two).  At separations $\gtrsim30 $ \kpc, the offsets maintain a roughly constant magnitude of $\sim$0.11 dex (a 30\% enhancement).   

\deloh~displays a much weaker trend with \rp~than the SFR, but similarly shows a sharp increase in offset magnitude at the smallest separations.  
Metallicity values are significantly suppressed by $-0.02$ dex out to $\sim$ 60 \kpc.  
The plateau seen in the SFR offsets is not apparent in the metallicity offsets for the total sample of pairs.
The relative weakness of this trend is not surprising, as metallicity shifts due to interactions are usually found to be a small magnitude effect, generally of order $-0.03$ to $-0.05$ dex \citep[e.g.,][]{Michel-Dansac2008, Cooper2008, Ellison2008, Ellison2009}.  

\subsection{Major \& Minor Mergers}
It is interesting to consider the difference in SFR trends for major versus minor mergers.  Our sample is large enough that we can divide according to the mass ratios of the interacting galaxies.  We note that we divide our galaxies by stellar mass instead of by magnitude \citep[cf.][]{Lambas2003, Woods2007, Lambas2012}; see \citet{Ellison2008} for a more extensive discussion of the use of luminosity, rather than mass, ratios.  Galaxies within a mass ratio $0.33 \leq M_{host} / M_{companion} \leq 3.0$ are considered major mergers, $M_{host} / M_{companion} > 3.0$ are more massive companions in a minor merger, and  $M_{host} / M_{companion} < 0.33$ are less massive companions in a minor merger.  Our sample contains 1116 galaxies in major mergers, and 783 in a minor merger, of which 184 are the more massive companion, and 599 are the less massive companions.  
The smaller number of more massive companions in minor mergers is likely due to the increased probability of high mass galaxies hosting an AGN \citep[e.g.,][]{Kauffmann2003} and thereby being removed from our star-forming sample.

\begin{figure}
\centerline{\rotatebox{0}{\resizebox{9cm}{!}
{\includegraphics{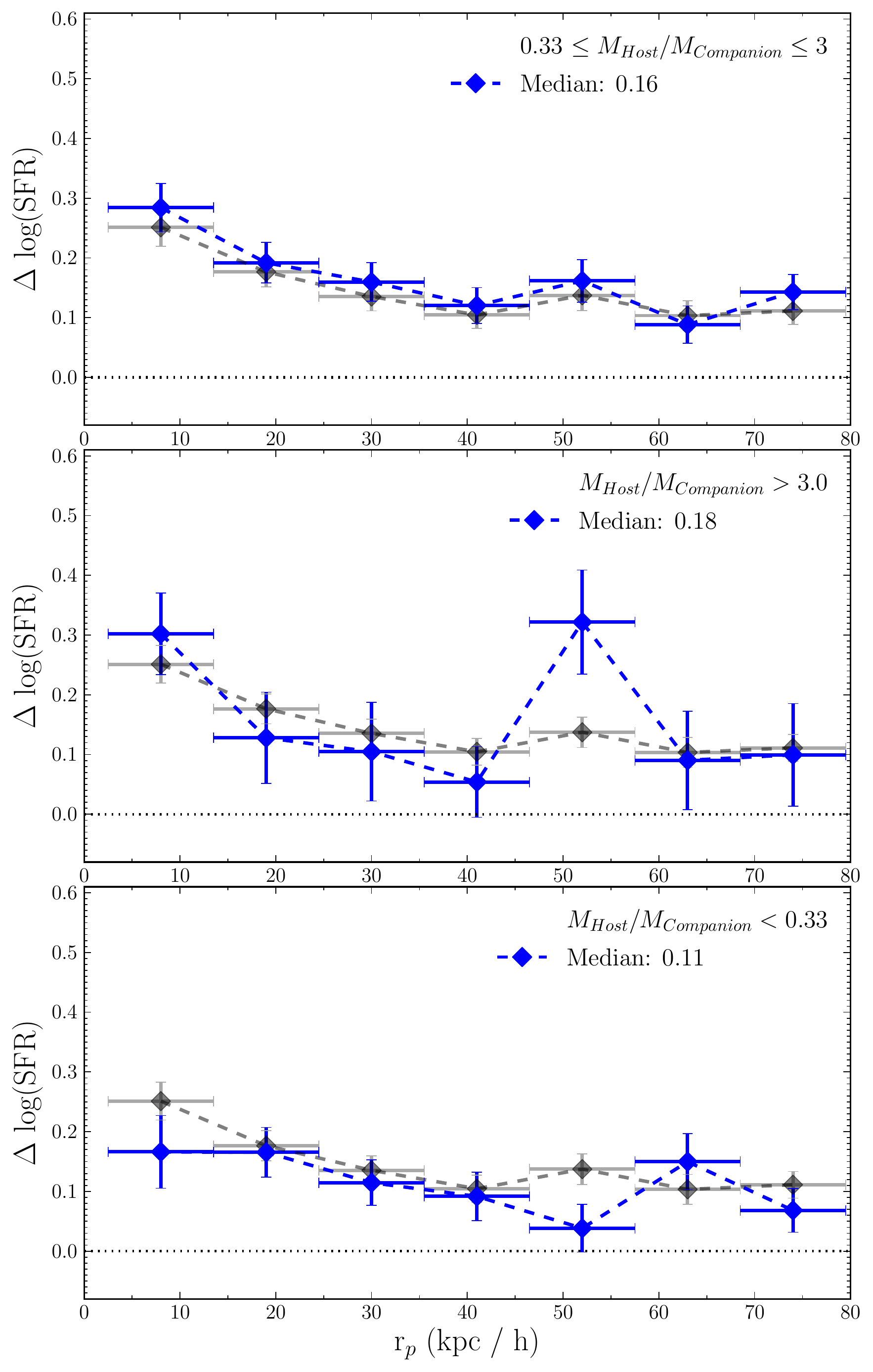}}}}
\caption{\label{fig:sfr_mratio} SFR offsets, separated by mass ratio.  Grey background points in all panels show the trends from the total sample (Figure \ref{fig:all_offset}).  The top panel shows the trend for all galaxies in major mergers (1116 galaxies).  Middle panel shows the more massive companion in a minor merger (184 galaxies) and the bottom panel shows the trend for the less massive companion in a minor merger (599 galaxies).  The more massive galaxies in a minor merger show an extremely strong enhancement at small separations, whereas the less massive companions show a much more consistent effect without any small \rp~increase in offset.  Both the more massive and less massive companions show similar levels of offset at wide separations ($\sim0.1$ dex).}
\end{figure}

The trends in \delsfr~in major, more massive companion, and less massive companion subsamples with \rp~are plotted in the top, middle, and bottom panels of Figure \ref{fig:sfr_mratio} respectively.  Grey background points are the SFR offsets from Figure \ref{fig:all_offset} in all panels.  Notably, galaxy pairs of all mass ratios are enhanced relative to the control at all separations.  
The shape of the offset vs. \rp~relation seen in Figure \ref{fig:all_offset} is apparently driven by the major mergers.  This is unsurprising, as major mergers are generally observed to show the strongest effects \citep[e.g.,][]{Woods2006, Woods2007, Ellison2008}, and they make up the majority (59\%) of our sample. 

With the exception of the more massive companion bin at 50 \kpc, both the more massive and less massive companions in a minor merger show relatively flat enhancements at the same magnitude ($\sim$ 25\% enhancement) at wider separations.  
However, the innermost bin shows a different response between more massive and less massive companions in a minor merger.  
The less massive companion shows no strong enhancement at small separations, distinguishing it from both the more massive companions and the major mergers, both of which show an increase in the magnitude of the SFR enhancement.  
More massive companions, on the other hand, show similar enhancement as the major mergers at small separations, at 2 times stronger than the control; major pairs are enhanced by a factor of 1.9.  
However, as the more massive companions have the smallest sample, each binned point in Figure \ref{fig:sfr_mratio} only has 20-30 galaxies.  While the use of a median means that we are not biased by one or two outlying points, the poor number statistics may result in an anomalously high point due to poor sampling of the total distribution.  A more detailed discussion of the offset distributions is presented in $\S$\ref{sec:offset_stats}.

As the significance of the metallicity offsets is much weaker than that of the SFR offsets, splitting the metallicity offsets into mass ratio bins does not provide any additional information.  

\subsection{Visual Classifications}
If the SFR enhancements and metallicity dilutions are truly being driven by the tidal interactions of galaxies in pairs, then selecting a subsample which shows morphological evidence of a recent tidal interaction ought to amplify the effects seen in Figure \ref{fig:all_offset} \citep[e.g.,][which found stronger effects in a morphologically disturbed subsample]{Michel-Dansac2008, Lambas2012}.  
To this end, we select only those galaxies which show strong tidal arms or other asymmetries induced by an interaction.  
Although the pairs sample is designed to minimize the inclusion of projected pairs, physically bound galaxy pairs which have not undergone their first pass should have SFRs and metallicities close to the control, and will weaken the interaction-triggered signal.
Any remaining projected pairs with low \delv~values would not be excluded from the sample, and, as physically dissociated systems, would also weaken the signal from the interacting systems.

\begin{figure}
\centerline{\rotatebox{0}{\resizebox{9cm}{!}
{\includegraphics{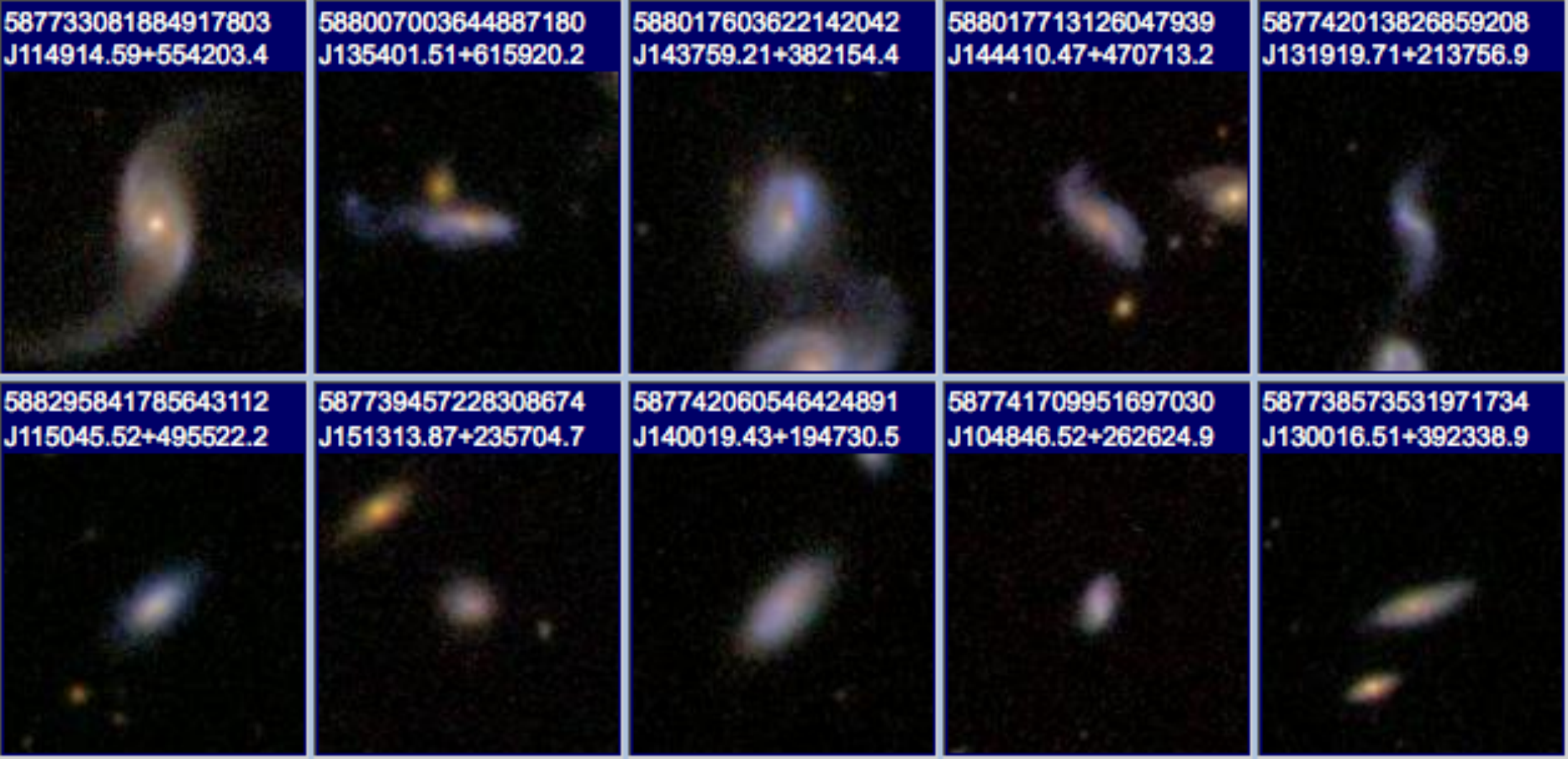}}}}
\caption{SDSS thumbnails of 10 randomly selected galaxies flagged as `visibly disturbed' (top row) or `not visibly disturbed' (bottom row).  All galaxies are labeled with their SDSS objids.}
\label{fig:mosaic}
\end{figure}

In order to eliminate the weakening effect due to either pre-interaction or projected pairs, all 1899 galaxies in the sample were visually classified for signs of morphological disturbances, as neither sample would be expected to show morphological signs of recent interactions.  Galaxies were flagged as either `visibly disturbed' or `not visibly disturbed'.  Galaxies which fall into the category of not being visibly disturbed will include pre-interaction galaxies and interacting galaxies whose tidal features are below the surface brightness limit of the SDSS imaging, along with galaxies which are not truly interacting,  so this category is not a useful diagnostic on its own.  However, the set of galaxies which falls into the disturbed galaxy classification should be a clean sample of galaxies which have already had a close encounter with a companion.  1105 galaxies are classified as disturbed, and the remaining 794 as `not visibly disturbed'  (See Figure \ref{fig:mosaic} for 5 randomly selected examples of each classification). 

\begin{figure}
\centerline{\rotatebox{0}{\resizebox{9cm}{!}
{\includegraphics{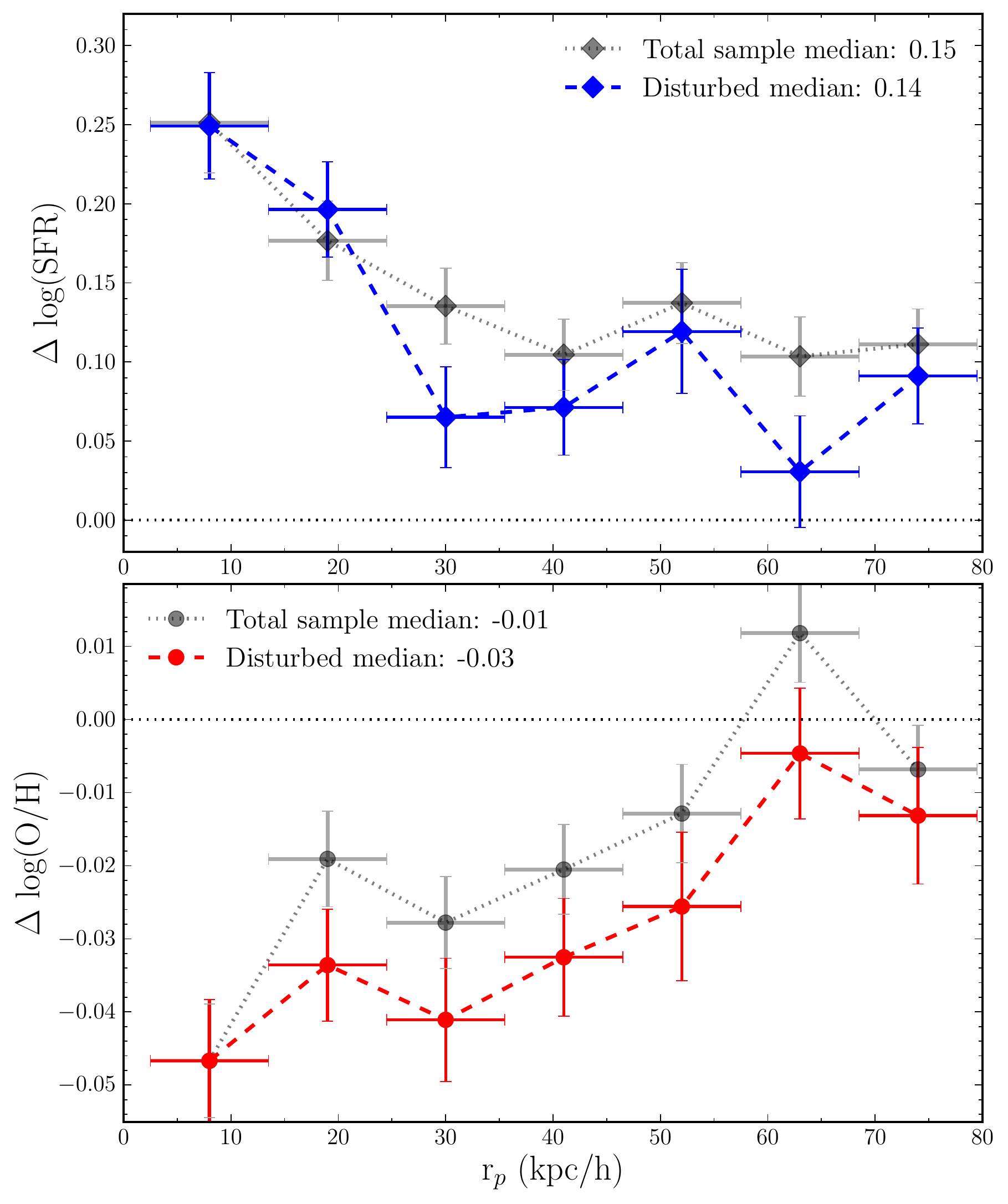}}}}
\caption{SFR and metallicity offsets for 1105 pair galaxies flagged as disturbed.  Grey background points are the trends from the total sample (i.e., Figure \ref{fig:all_offset}).  SFR offsets are of approximately equal magnitude in the disturbed and total samples, but the metallicity offsets are significantly larger in magnitude for the disturbed sample than for the total sample.  The median value of \deloh~triples in magnitude when the disturbed sample only is taken.}
\label{fig:anydisturbed}
\end{figure}

The SFR and metallicity trends with \rp~for the subsample of visibly disturbed galaxies are shown in Figure \ref{fig:anydisturbed}.  Grey points are the total sample points from Figure \ref{fig:all_offset}; the disturbed subsample is overplotted in coloured points.   
The SFRs do not show a systematic boost to their offsets in the morphologically disturbed sample; most points in the disturbed subsample are consistent with, or slightly below, the total SFR offsets.
A KS test reveals an 89.78\% chance that the disturbed and total samples were drawn from the same parent population.
However, the median metallicity has shifted to more metal-poor values in all \rp~bins, in some cases by nearly $-$0.02 dex.  The median \deloh~has tripled in magnitude to $-0.03$ dex for the disturbed sample.  The disturbed and total metallicity samples have a KS probability of being drawn from the same distribution of only 1.29\%.  Within this subsample, galaxies are significantly metal poor out to 80 \kpc, the same range over which the SFRs show enhancements.  

The fact that the SFRs show no significant enhancement in the disturbed subsample is a puzzling contrast to the significantly lower metallicities.
To ensure that the shifts between the total sample and the disturbed subsample are not simply due to the smaller size of the disturbed subsample, we bootstrap a random sample of 1105 galaxies from the full pairs sample, and find the median value of the random subsample in both SFR and metallicity.  This resampling was then repeated 20,000 times.  We find that it is very easy to obtain a SFR offset of +0.14 dex through random sampling of the \delsfr~distribution, but extremely difficult to reach median \deloh~values of --0.03 dex at random.  
The bootstrap test indicates that the lower metallicity in the disturbed subsample is unlikely to be a statistical fluke due to a smaller sample size.  However, the median SFR offset in the disturbed subsample is consistent with the total sample, and no significant difference in the SFR offsets of the disturbed sample is seen.

One potential interpretation is that the timescales over which galaxies are visibly disturbed are more strongly correlated with the timescales in which metallicity dilution is the strongest, and only weakly correlated with strong star formation enhancements.  
There is some suggestion from simulations that tidal features are visible for both a shorter period of time than the SFR burst, and at earlier times than the SFR enhancement \citep{Lotz2008}.  If this is the case, then a sample of galaxies selected to be morphologically disturbed would not necessarily be expected to identify the galaxies with the largest \delsfr.

\section{Distributions of SFR and metallicity offsets}
\label{sec:stats}
While the median offset values as a function of \rp~prove useful as a metric of the typical merger, it is interesting to explore the range of metallicity and SFR offsets.  The distributions will provide, for example, insight into how often the most extreme offsets occur, as well as how frequently no statistical change is seen.
Furthermore, since simulations suggest that the two galaxies involved in a major merger ought to show similarly enhanced SFRs \citep{Montuori2010, Torrey2012}, we can search for observational evidence of simultaneous triggering on a merger-by-merger basis.

\subsection{Offset distributions: Pairs vs. Controls}
\label{sec:offset_stats}
\begin{figure}
\centerline{\rotatebox{0}{\resizebox{9cm}{!}
{\includegraphics{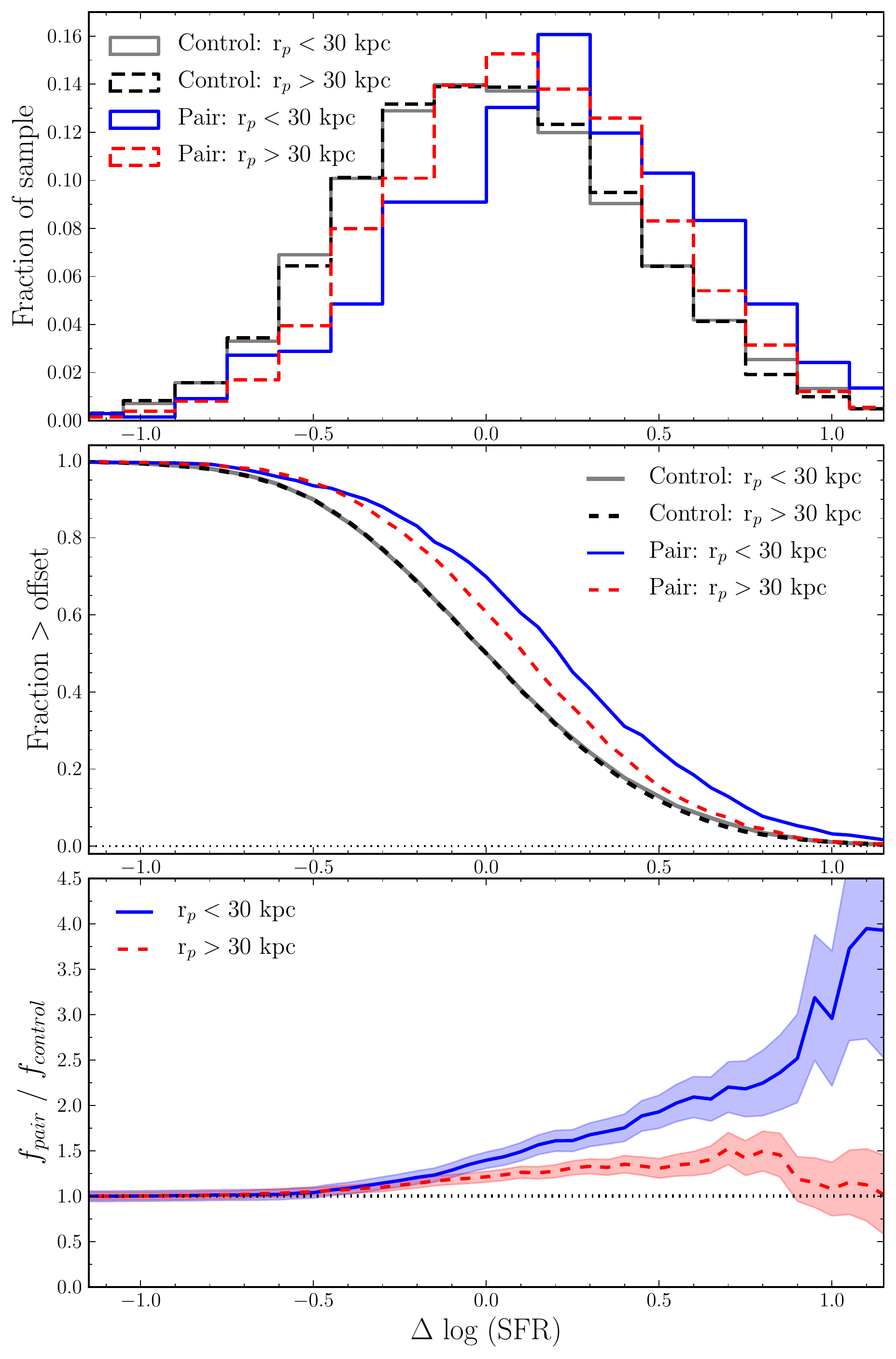}}}}
\caption{\label{fig:sfr_enhance} Top panel: histogram of SFR offsets for close pairs (blue solid), wide pairs (red dashed), and their controls.  Control offsets are calculated between all controls matched to the same galaxy.  Both close and wide pair galaxies are visibly shifted to higher offset values than the controls.  Middle panel: cumulative distribution of SFR offsets, indicating the fraction of the galaxies in the sample with offsets greater than a given value.  Galaxies are divided in the same way as the top panel.  At all offset values, the close pairs sample shows a higher fraction of galaxies with strong SFR enhancement.  The bottom panel shows the ratio of the fraction of pair galaxies above the threshold offset to the fraction of the control above that offset, with the shaded region indicating $\sqrt N$ errors. Black dotted line shows a ratio of 1.}
\end{figure}

\begin{figure}
\centerline{\rotatebox{0}{\resizebox{9cm}{!}
{\includegraphics{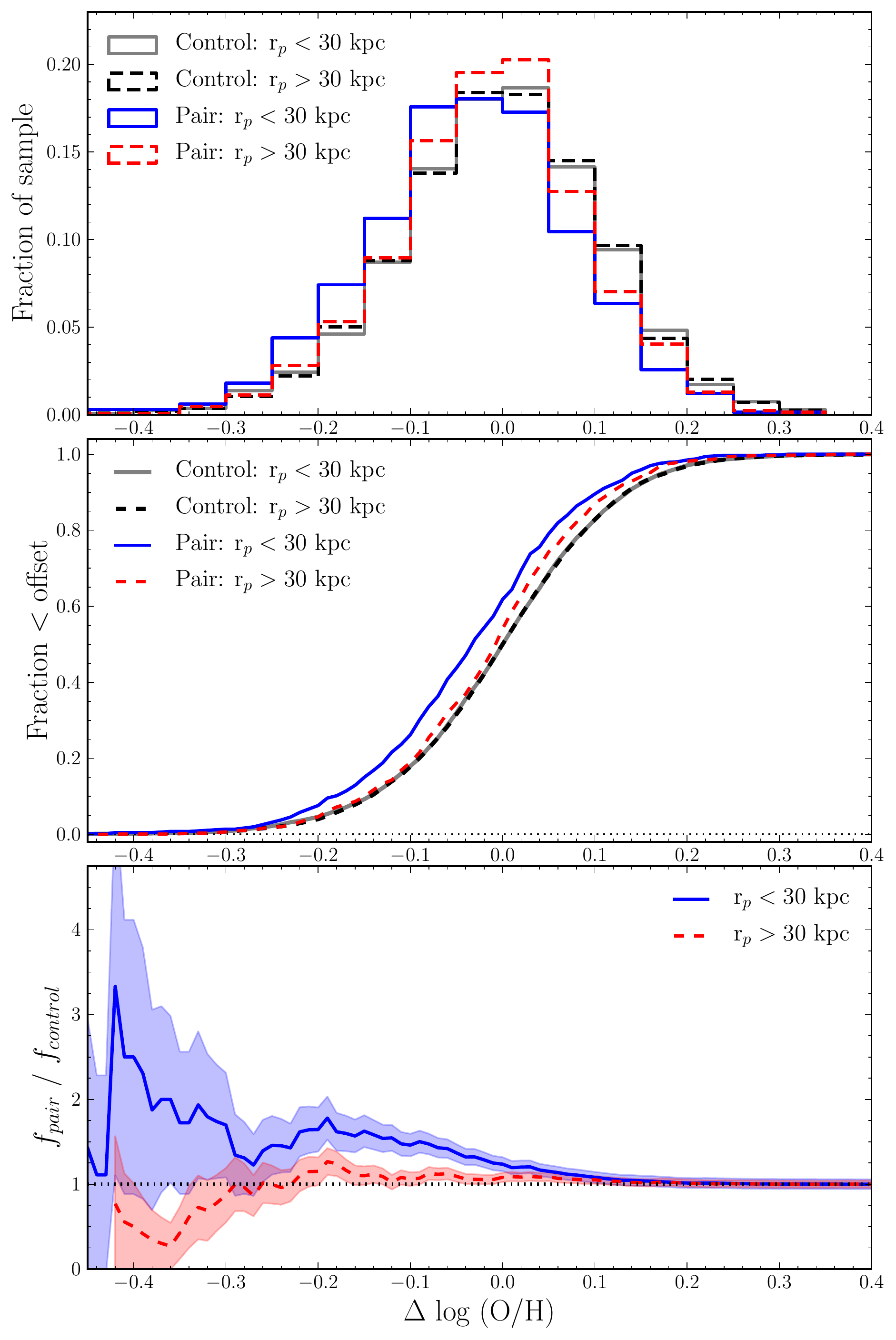}}}}
\caption{\label{fig:mzr_enhance} Same as Figure \ref{fig:sfr_enhance}, but for metallicities.  The top panel shows the distribution of the pairs and control samples, split into \rp\ $<30$ kpc and \rp\ $>30$ kpc bins, coloured same as Figure \ref{fig:sfr_enhance}.  Middle panel shows the cumulative distribution of metallicity offsets, and bottom panel shows the ratio of the pair and control fractions for $>30$ and $<30$ kpc bins. The shaded region indicates $\sqrt N$ errors.}
\end{figure}

The pairs sample is first split into two bins of \rp. We define all galaxies with \rp~$<30$ \kpc~as `close pairs', and any galaxy pairs with \rp~$>30$ \kpc~as `wide pairs'.  The distributions of \delsfr~and \deloh~are shown in Figures  \ref{fig:sfr_enhance} \& \ref{fig:mzr_enhance} respectively, with blue solid lines for the close pairs sample and red dashed lines for the wide pairs.  
The control sample is also divided by taking all control galaxies matched to close pairs as the \rp~$<30$ \kpc~control sample, and all galaxies matched to wide pairs as the \rp~$ >30$ \kpc~control sample. 
Control offsets were calculated in a similar way as the pair offsets.  
Each galaxy in the control sample was compared to the set of other galaxies matched to the same pair galaxy.  As control galaxies were matched in sets of ten, every control galaxy has 9 galaxies of similar mass, redshift, and local density, to which it can be compared.  
We therefore take the median of the 9 other control galaxies, and find the difference between that median value and the control galaxy selected.  
This difference (analogous to the calculations in Equation \ref{eq:mzr_offset}) produces a control offset for every control galaxy in both SFR and metallicity.  
As the control galaxies are tightly matched, the control offsets ought to be centred around zero offset in both SFR and metallicity, with no significant shift away from zero or between the close and wide control samples.  Figures \ref{fig:sfr_enhance} \& \ref{fig:mzr_enhance} show the control samples as the solid grey (close controls) and dashed black (wide controls) lines. The two control samples trace each other extremely well, and are centred around zero.

The top and middle panels of Figures \ref{fig:sfr_enhance}  \& \ref{fig:mzr_enhance} offer two ways of viewing the distribution of offsets for SFR and metallicity.  The top panel shows the offsets in a discrete way, with the overall form of the offset distributions for the two subsamples of both pairs and controls, normalized to the sample size.  The middle panel shows the cumulative distribution of the same samples, which illustrates the differences between the distributions more clearly.
Figures \ref{fig:sfr_enhance}  \& \ref{fig:mzr_enhance} show the total pairs sample rather than the disturbed subsample in order to maintain the number statistics of our sample.  While the full sample will contain a small fraction of galaxies which are not physically associated, and a larger fraction of galaxies which have not yet interacted, this should only increase the number of galaxies with offsets near zero, and will not affect the distribution of offsets further from zero.

In the top panel of Figure \ref{fig:sfr_enhance}, it is already clear that both the close and wide pair samples are shifted systematically to higher offsets than their respective control samples, a trend which is even more dramatic in the middle panel.  The cumulative distribution is calculated as the fraction of the total sample which has an offset greater than a given threshold.  As the threshold values increase, the fraction of the control samples which have strong offsets decreases; the pairs decline more slowly.  KS tests confirm this visual offset; both close and wide pairs are inconsistent with being drawn from the same parent population as their controls at $>6\sigma$. To quantify this in a different way, the median \delsfr~of the close pairs sample is $+0.21$ dex (or a 60\% enhancement over the control; see also Figure \ref{fig:all_offset}), whereas the wide pairs sample has a median \delsfr~of $+0.11$ dex (a 30\% enhancement).

The bottom panels of Figure \ref{fig:sfr_enhance} shows the ratio of the pair fraction greater than a given offset (seen in the centre panel of Figure \ref{fig:sfr_enhance}) relative to its respective control fraction, as a function of offset value.  The black dotted line indicates a ratio of 1:1, i.e., that the pair and control samples have exactly the same fraction of the total sample at that offset or greater.  
This panel illustrates the relative frequency with which a given offset will appear in the pairs sample instead of the control.  
Values higher than one indicate that the offset is preferentially found in the pairs sample.  The shaded regions indicate the range of $1\sigma$ errors, calculated from $\sqrt{N}$ statistics.
This panel indicates that as \delsfr~becomes more extreme, it is increasingly likely to observe these offsets in the close pairs sample, rather than the controls or the wide pairs sample.  The wide pairs sample also shows an excess of positive \delsfr~relative to the control sample, although only up to offsets less than  +0.85 dex (7 times the control value) 
Therefore, while the median SFR offset is only an increase of 40\%, there is an excess of close pairs with SFR enhancements up to a factor of 10 stronger than the control.  Although only $\sim3$\% of pairs have excesses at this level, this is 3 times more than exist in the control.  Wide pairs, by contrast, only show excesses up to a factor of 7 (0.85 dex).

Figure \ref{fig:mzr_enhance} is set up in the same way as Figure \ref{fig:sfr_enhance}.  The systematic shift of the metallicities in the pairs sample is less pronounced than that of the SFR offsets, but is still visible.  KS tests also verify this visual trend, with the metallicities for both close and wide pairs being inconsistent with being drawn from their parent distribution at $>4\sigma$.  Here the cumulative distribution is calculated as the fraction of galaxies with offsets \textit{less} than the tested offset.  The close pairs show a distinct shift to lower values relative to the control, whereas the wide pairs do not show a particularly strong shift.  The difference in median is also much weaker; close pairs have a median offset of $-0.034$ dex (8\%), and the median for the wide pairs is $-0.01$ dex (2\%).

This distinction between close and wide pairs in the metallicities is confirmed in the bottom panel, which shows the relative frequency of a given offset between the pair and control samples.  Here we see that the the wide pairs do not show a significant excess of metallicity offsets relative to the control at any magnitude (see also Figure \ref{fig:all_offset}).  However, the close pairs show a significant excess of metal-poor galaxies at all negative offsets up to $-0.25$ dex  (up to 1.78 times lower than the control).  Offsets more extreme than $-0.25$ dex are so rare that the significance is overwhelmed by poor number statistics.  The excess of very low metallicity galaxies is interesting, considering that the median offset for the pairs is $\sim-0.02$ dex, or 5\% lower than the control.  The discrepancy between the magnitudes of the excess offsets in the pairs sample and the overall median seems to indicate that the median offset is at least partially driven by a small number of galaxies which are very strongly offset from the controls.

\begin{figure}
\centerline{\rotatebox{0}{\resizebox{9cm}{!}
{\includegraphics{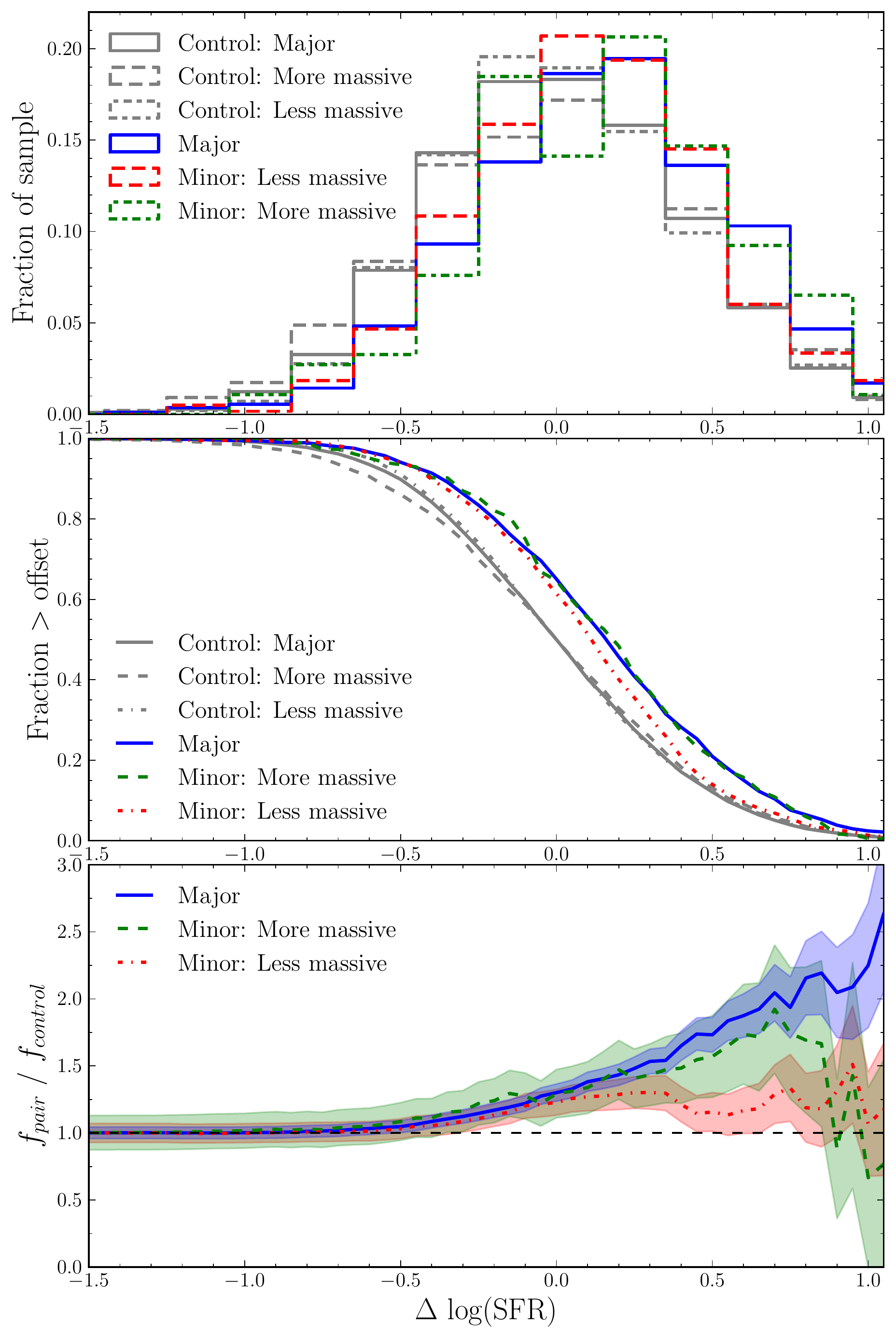}}}}
\caption{\label{fig:mratio_dists} Same as Figure \ref{fig:sfr_enhance}, but split by mass ratio instead of \rp.  The top panel shows the distribution of offsets for the three mass ratio bins: major mergers ($0.33\leq$ M$_{Host}$/M$_{Companion} \leq3$), the less massive companion in a minor merger (M$_{Host}$/M$_{Companion}<0.33$), and the more massive galaxy in a minor merger (M$_{Host}$/M$_{Companion}>3$).  The control galaxies for the galaxies in each mass ratio bin are plotted in grey. The centre panel shows the fraction of each of the three samples (and respective controls in grey) which has an offset greater than a given value.  The bottom panel shows the ratio fraction of the pairs galaxies above the threshold offset to the fraction of the control above that offset.  Shaded regions indicate $\sqrt{N}$ errors.}
\end{figure}

If, instead of splitting the total pairs sample into close pairs and wide pairs, we divide the sample into bins of mass ratio, we can investigate how the offset distributions depend on the mass ratio of the merger.  The pairs sample is divided into major mergers ($0.33\leq$ M$_{Host}$/M$_{Companion} \leq3$), less massive galaxies in a minor merger (M$_{Host}$/M$_{Companion}<0.33$), and the more massive galaxies in a minor merger (M$_{Host}$/M$_{Companion}>3$).  The control samples are split so that the control galaxies matched to pair galaxies in each bin in mass ratio are assigned to their respective bins.  
Figure \ref{fig:mratio_dists} is constructed identically to Figures \ref{fig:sfr_enhance} \& \ref{fig:mzr_enhance}.
The top panel of Figure \ref{fig:mratio_dists} shows that all three mass ratio bins (in colour) are systematically shifted to higher SFR values than their control samples (in grey).  The control samples overlay each other reasonably well and are centred around zero, as expected.  Given the similarity of the median values in Figure \ref{fig:sfr_mratio}, it is unsurprising to see that the 3 distributions are shifted by approximately the same amount relative to the control samples.  KS tests give probabilities indicating that none of the galaxy pair subsamples are likely to be drawn from the same parent distribution as their respective control samples at $>5\sigma$ in all cases.
The middle panel of Figure \ref{fig:mratio_dists} shows the fraction of each mass ratio subsample which has an offset greater than a given threshold, as a function of that threshold.  Here again we can see that the three distributions are roughly consistent in their shift to higher \delsfr, relative to the control sample. 
KS-tests on the distributions of pair offsets in the 3 mass ratio subsets indicate that the 3 distributions are consistent with being drawn from the same parent population; none of them are inconsistent with the null hypothesis at $>3\sigma$ confidence.  

The similarity between the offset distributions between mass ratio bins is reinforced in the bottom panel of Figure \ref{fig:mratio_dists}.  
All three samples show an excess of positive values of \delsfr~relative to the control.
Both major and minor mergers seem equally effective at inducing offsets up to 0.45 dex above the control value, or an enhancement of a factor of $\sim3$.  These intermediate offsets are roughly 1.3 times as likely to occur in the pairs sample as the control in all samples, with no significant distinction between the more massive and less massive companions in the minor mergers, or between the minor mergers and the major mergers.  SFR enhancements of a factor of 2 (+0.3 dex) occur in 30-35\% of the pairs sample, and are 1.35--1.65 times more likely to occur in the pairs than in the control. 
However, offsets more extreme than 0.8 dex are reached almost exclusively through major mergers.  These offsets are just as rare as before, occurring in roughly 5\% of galaxies, but are more than twice as likely to occur in major pairs as in the control.

In summary, extreme offsets are rare in the pairs sample, but are most likely to be found in the close pairs with approximately equal masses.  SFR enhancements of +1.0 dex (a factor of 10 higher than the control) are 3 times as likely to occur in the pairs as in the control, but occur in only 3\% of the total pairs sample.  Similarly, metallicity offsets of --0.25 dex (a factor of 1.78 lower than the control) are 1.5 times as likely to occur in the pairs as in the control, but also are only found in 3\% of the pairs sample.   Conversely, modest SFR enhancements, up to a factor of $\sim$ 3 beyond the control, can be found as readily in the minor mergers as the equal mass pairings.

\subsection{Evidence for synchronised SFR triggering}
Previous work has found that galaxies in pairs often have both galaxies show enhanced star formation more frequently than would be expected at random \citep{Xu2010}.  This is interpreted as the signature of the two galaxies in an interaction undergoing synchronised SFR enhancement, as both galaxies undergo similar tidal torques due to their companion.  This physical picture is supported by simulations of equal mass mergers, where both galaxies tend to show similar responses to the interaction \citep[e.g.,][]{Torrey2012}.  Some observational evidence for merger-driven synchronicity between galaxies has already been found in galaxy pairs with AGN.  \citet{Ellison2011b} found that in galaxy pairs at small separations, a galaxy hosting an AGN was twice as likely to have a companion also hosting an AGN than would be expected at random. We therefore wish to investigate whether the \delsfr~values also show evidence of synchronised enhancements.

\begin{figure}
\centerline{\rotatebox{0}{\resizebox{9cm}{!}
{\includegraphics{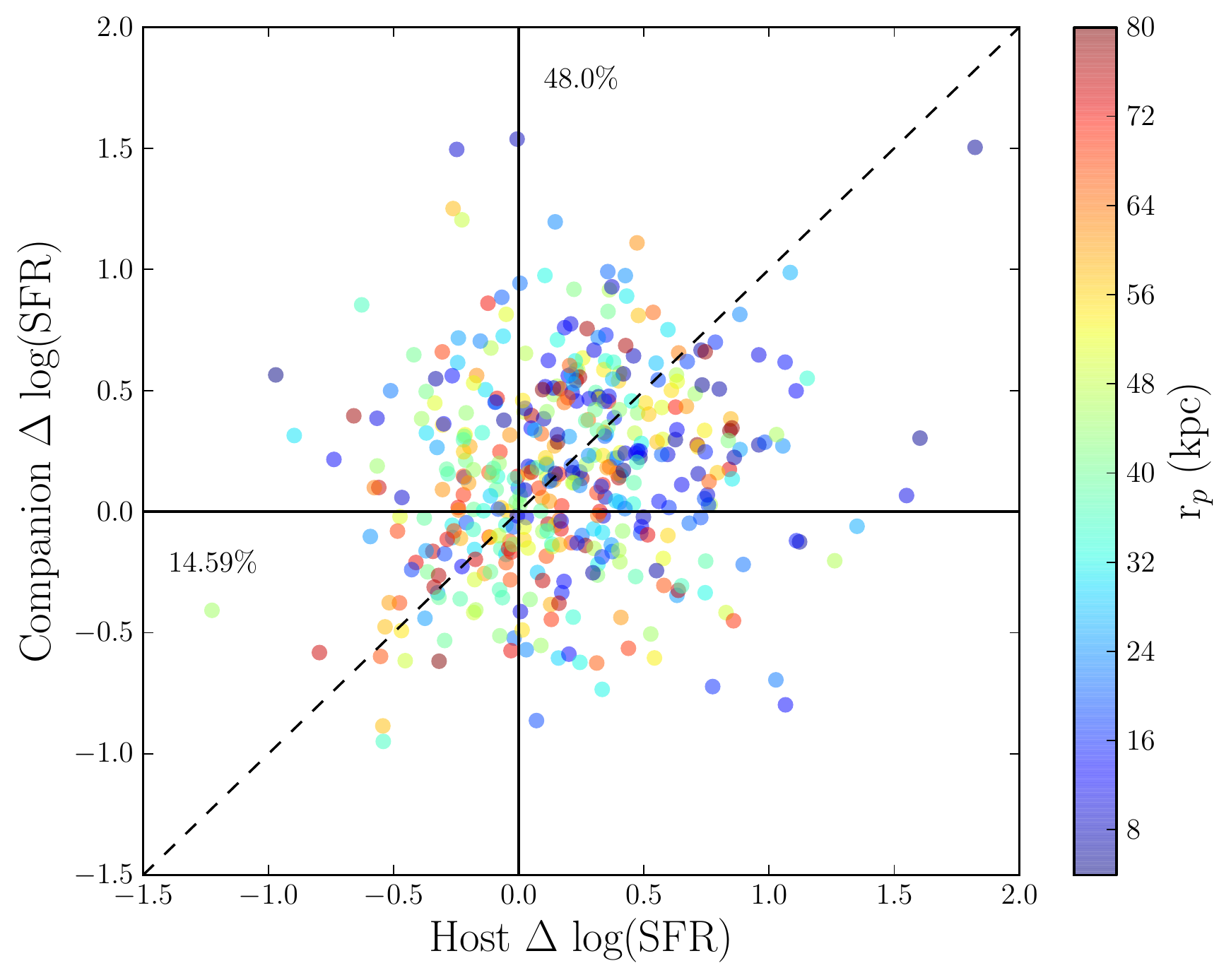}}}}
\caption{\label{fig:sfr_trigger} For the subset of galaxy pairs where both the host and the companion are in our sample, ($\sim$45\% of the sample; 425 galaxy pairs), the distribution of the galaxy pair \delsfr~vs. its companion's \delsfr.   The colour bar indicates the projected separations of the pairs.  48\% of the sample is found where both the pair and its companion have \delsfr~$>$ 0 (top right quadrant).  Using a random galaxy pair match of control sample galaxies, we find each quadrant to be equally populated.  Galaxies in pairs are 1.9 times as likely to show double SFR enhancements than a random control sample, and 0.6 times as likely to show double suppressions (bottom left quadrant).  Relative to a false pairs sample, the galaxies in true pairs are $>$17\% more likely to show double enhancements.}
\end{figure}

We first select only those galaxies from the total pairs sample where both the pair and the companion are found in our sample.  This reduces us to 45\% of our original sample; 425 galaxy pairs remain in the sample (850 galaxies).  We then plot the host \delsfr~versus its companion's \delsfr~(Figure \ref{fig:sfr_trigger}), where the points are colour-coded according to projected separation.  Each galaxy pair is plotted only once on this diagram.  It is clear that there is an overabundance of points in the double-enhancement (top right) quadrant of the figure; this quadrant contains 48.0\% of the total pairs sample.  The double-deficit quadrant contains only 14.59\% of the sample.  To gain a sense of what would be expected at random, we pair random control galaxies with each other, and find the distribution of points for the control offsets calculated in Section \ref{sec:stats}.  As expected, each quadrant contains 25\% of the control pair distribution.  This indicates that the pairs are 1.9 times as likely to fall in the double-enhanced quadrant than would be expected from the control.  Splitting this sample into galaxies in major mergers and those in minor mergers results in an almost identical fraction of galaxies with doubly enhanced SFRs (47.42\% in major mergers vs 49.25\% in minor mergers).  The minor mergers show a slightly lower fraction of galaxies with double suppressions (2\%) relative to the major mergers.

However, since the \delsfr~values are higher in the pairs sample than they are in the control sample, simply having an excess of galaxies with doubly enhanced SFRs does not necessarily indicate that there is synchronous triggering.  This could instead be the result of comparing between two samples which are not drawn from the same parent distribution, rather than indicating that pairs of galaxies are showing correlated SFR enhancement.  In order to eliminate the issue of the differing parent distributions, instead of comparing the population of the pairs in this diagram to the control sample, we compare to the pairs sample itself.  We reassign each galaxy in the sample to a random companion, thereby scrambling the pairs sample.  This results in a set of uncorrelated pair \delsfr s which can be plotted on the same diagram, but which will have the same distribution of offsets as our true pairs sample.  We find that the scrambled pairs populate the double enhancement quadrant with 40.99\% of the total sample.  13.28\% of the scrambled pairs sample falls in the double-suppression quadrant, identical to the true pairs sample.  However, the double enhancement quadrant has 17\% more galaxies in the true galaxy pair sample than in the scrambled pairs.  

The SFR offsets have some intrinsic scatter, and galaxies very close to the zero line may simply be scattered to one side of zero or another.  To minimize the effect of this scatter, we can include a buffer around the zero value, such that galaxies within a certain range around the 0 offset value are not counted in this fractional counting.  If the buffer is used, the strength of the double SFR enhancement increases.  If galaxies which have offsets beyond $\pm$0.13 dex (the 25th percentile for the control distribution) are considered in the fractional calculation, then the true pairs are 34\% more likely to show double enhancements.  

\section{Comparison with Theoretical Models}
\label{sec:sims}
The SFRs and metallicities of our pairs are significantly offset from the control over significantly larger distance scales than has been previously seen \citep[e.g.,][]{Lambas2003, Alonso2006, Woods2007, Ellison2008}.  The interpretation of the smoothly declining trend with increasing \rp~has traditionally been that galaxies promptly undergo a burst of star formation as the galaxies reach pericentre.  This burst would then decline to a fiducial value as the galaxies separate.   However, Figure \ref{fig:anydisturbed} indicates that galaxies are still offset from the control values at $\sim$80 \kpc~separations, and the wider separations show a plateau instead of a smooth decline to control values.   The existing interpretation is insufficient to account for the wide separation plateau, as even a long lasting starburst should still show a smooth decline to the control values.
We therefore turn to theoretical models to aid in the interpretation of these trends.

We make use of simulations of major galaxy mergers developed in \citet{Torrey2012} to help interpret our results.  A detailed analysis of varying mass ratios and orbital parameters has previously been explored elsewhere \citep[e.g.,][]{diMatteo2007, Cox2008}.
Here, the goal is to take a general look at the SFR and metallicity changes in a merger as a function of \rp.  We refer the reader to \citet{Torrey2012}  for a detailed description of the simulation setup and parameters.  Briefly, \citet{Torrey2012} presents a set of N-body/Smooth Particle Hydrodynamics simulations using {\sc{Gadget-2}} \citep{Springel2005a}.  These models include cooling, star formation, feedback, and chemical enrichment.  Within these simulations, the metallicities are defined as the mass-weighted average of the metallicities of all gas particles within a sphere of 1 kpc around the centre of the galaxy.  Increasing the radius of this sphere only mildly alters the results of the simulations, and does not affect our conclusions.  (As a comparison, the physical diameter of the SDSS fibre is generally of order of a few \kpc.)  Galaxies are shown to be stable (i.e., do not develop a bar) for at least 2 Gyr when modelled in isolation.  For the purposes of constructing an interpretive framework for the SDSS data presented here, we use a suite of 16 galaxy mergers, varying only the galaxy orientations between mergers.  As a result, their orbits are kept constant through the suite.  The model galaxies are merged with identical copies of themselves in all 16 simulations \footnote{Our model galaxy is Disk B in Table 1 of \citet{Torrey2012}, but with a 25\% gas fraction, and merged on all 16 orientations in Table 2 of \citet{Torrey2012}.}.  

\begin{figure*}
\centerline{\rotatebox{0}{\resizebox{13cm}{!}
{\includegraphics{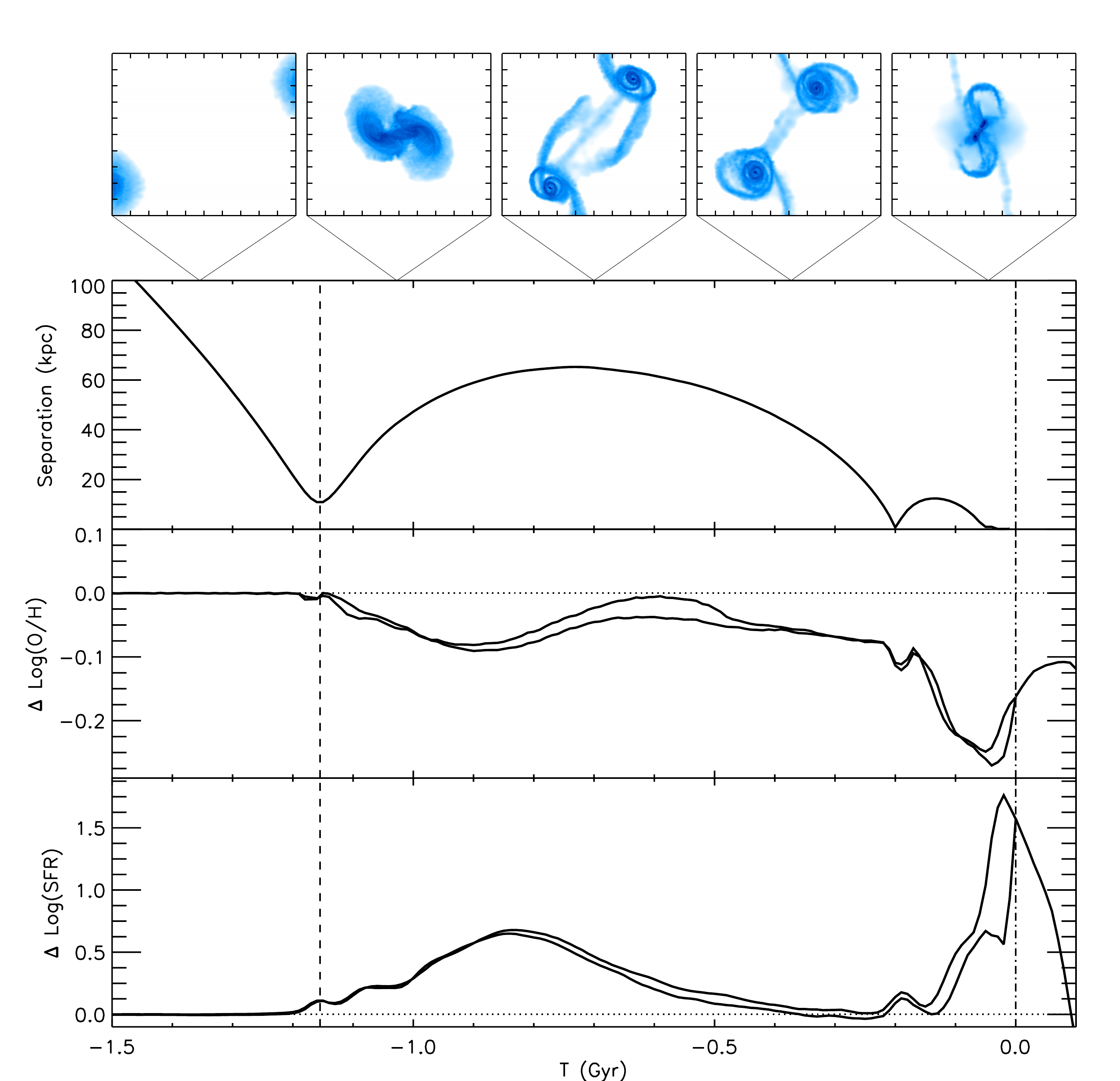}}}}
\caption{Simulation snapshots, \delsfr~(within the central 1 kpc), \& \deloh~as a function of time.  The dashed line marks first pericentre, and the dot-dashed line indicates coalescence. The top series of panels shows the surface density of the gas within the galaxy at different points in the merger.  \deloh~is the difference between the metallicity of a quiescent disk and the metallicity of the merging galaxy at any given time step, and \delsfr~is the difference between the current SFR at a given time and the SFR of a galaxy modelled in isolation.  The two solid lines in the metallicity and SFR panels indicate the responses of the two galaxies in the simulation.  The separation panel shows the intergalactic distance as a function of time, which remains fixed through the full suite of simulations; \deloh~and\delsfr~change between simulations as the galaxy orientations change.}
\label{fig:sim_timeline}
\end{figure*}

An example of the evolution of one of the 16 mergers is shown in Figure \ref{fig:sim_timeline}.  This figure shows the time evolution of the separation of the two nuclei in the top panel, the change in metallicity in the centre panel, and the change in SFR for both galaxies in the bottom panel, where $t=0$ is scaled to coalescence.  Snapshots of the galaxies' gas densities are shown as insets at the top of the figure.  Comparable values to the observational \deloh~and \delsfr~values are extracted from the simulations.  \deloh~and  \delsfr~are calculated as the difference between the merging galaxy's SFR or metallicity at any given time and the SFR or metallicity of a model quiescent disk at that same time.  The two lines in the bottom two panels show the responses of the two galaxies in the merger.  For major mergers, the close tracking of the two lines is a general feature of the merger simulations.  The magnitudes of the SFR enhancement and metallicity dilution may shift slightly from merger to merger, but the overall shape of these tracks with time is consistent across the suite of simulations.   Therefore, in some runs, the metallicities may re-enrich past the initial value, but this re-enrichment is almost always bracketed by periods of metallicity dilution.  Similarly, the strength of the initial SFR burst may vary slightly from merger to merger, but always diminishes again, and is dwarfed in magnitude by the peak in SFR as the galaxies reach the end of their merger. 

Of particular note in Figure \ref{fig:sim_timeline} is the time delay between first passage and the strongest metallicity dilutions and highest SFR enhancements after pericentric passage.  Metallicity dilutions do not reach their strongest values until $\sim$ 200 Myr after first passage; the peak in SFR enhancement is slightly longer, at $\sim$ 275 Myr\footnote{These timescales are primarily set by the free-fall timescales of the systems, given a rotational velocity of the simulated galaxy of 130 km/s.  }.  This delay means that the galaxies have had time to progress out to wider separations by the time the triggered SFR enhancement or metallicity dilution will be strongest.  There also is a period of metallicity re-enrichment after the first starburst, prior to coalescence.  Recently formed stars will return enriched gas to the interstellar medium as they reach the end of their lifespans, which drives the galaxies to nearly return to their initial metallicity.
The flow of metal-poor gas to the central regions of the galaxies continues throughout the interaction, re-diluting the central metallicities.   As galaxies progress into final coalescence, metallicity offsets drop to their lowest values, and \delsfr~reaches the strongest enhancement of the merger sequence, due to the large torques exerted as the nuclei coalesce. 

\begin{figure*}
  \centering
  \begin{minipage}[c]{8 cm}
   \includegraphics[width=220px]{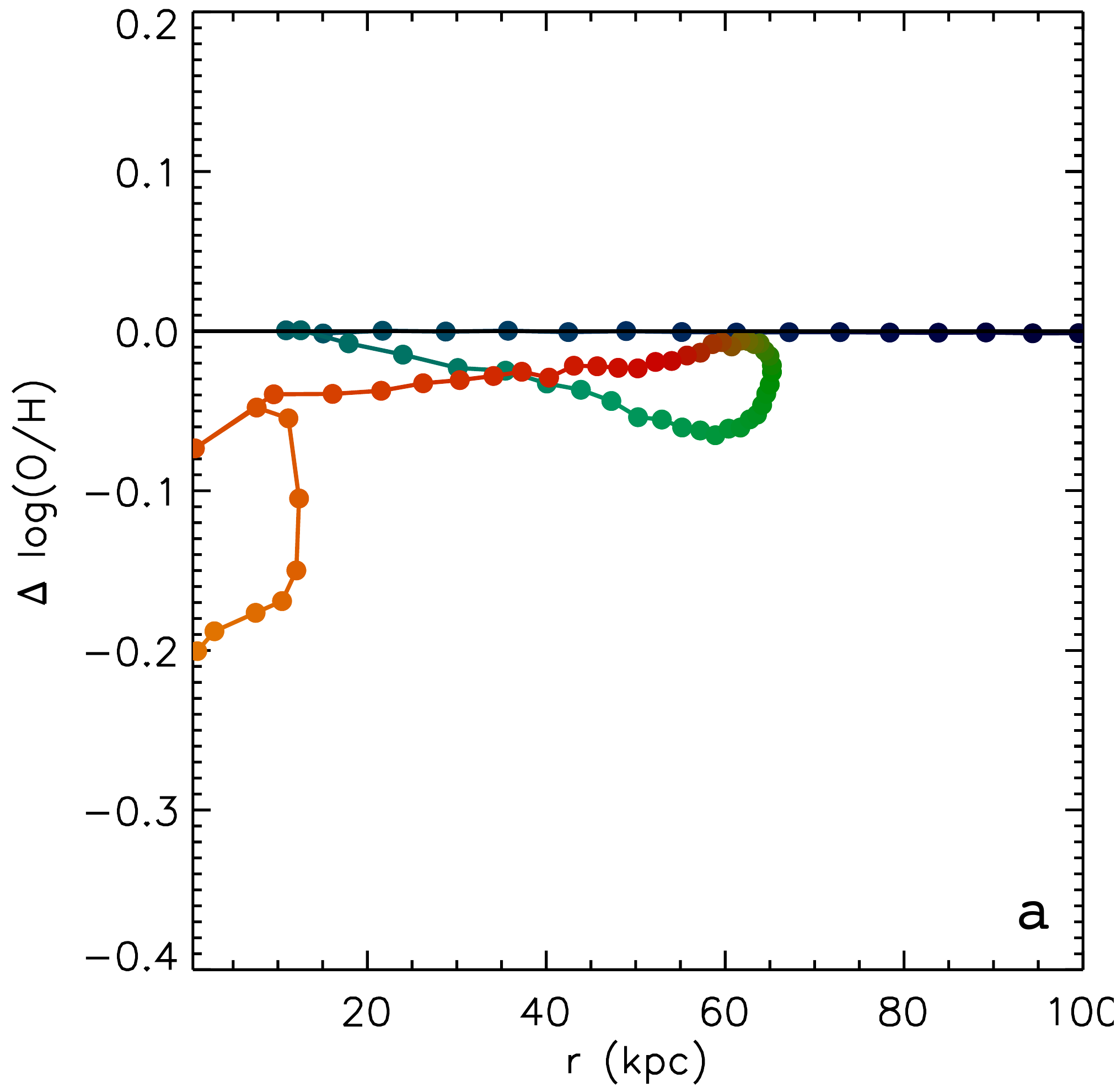}
   \label{fig:met_track}
  \end{minipage}
  \begin{minipage}[c]{8 cm}
    \includegraphics[width=220px]{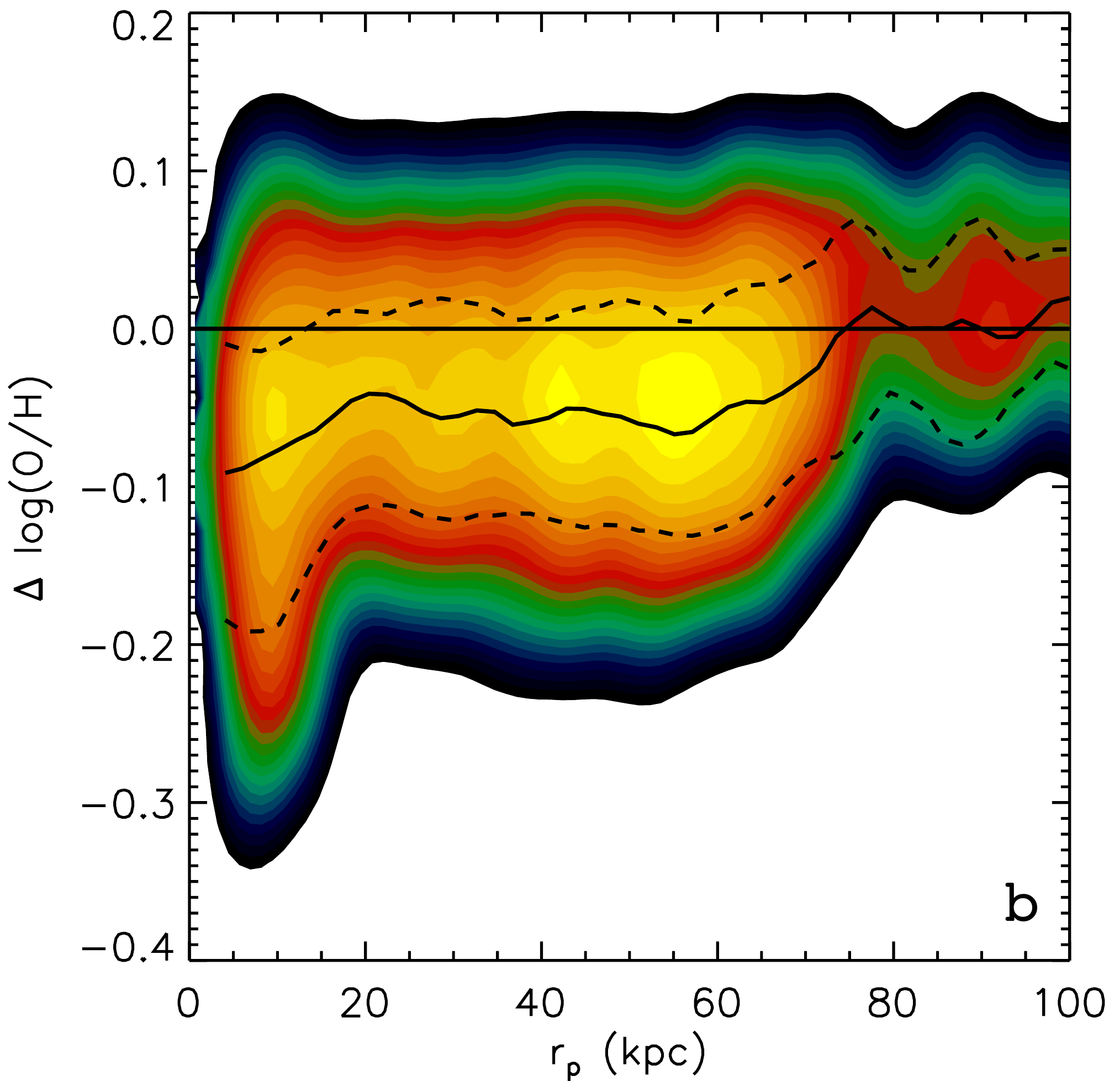} 
  \end{minipage}
  \caption{Panel a:  1 sample track of the metallicity offset as a function of true separation.  Colour indicates the progression of time through the merger, with dark blue at early times to yellow at late times.  Each dot indicates a fixed time step of 10 Myrs as the simulation progresses.  Panel b: 16 tracks, converted into projected separations.  Colours indicate the density within the contours, with yellow indicating the highest density of points from the simulation and blue indicating the lowest density.  The solid black line overlaid on the contours indicates the median value, and the dashed black lines indicate the 25$^{th}$ and 75$^{th}$ percentile range.}
\label{fig:met_sims} 
\end{figure*}

\begin{figure*}
  \centering
  \begin{minipage}[c]{8 cm}
\includegraphics[width=220px]{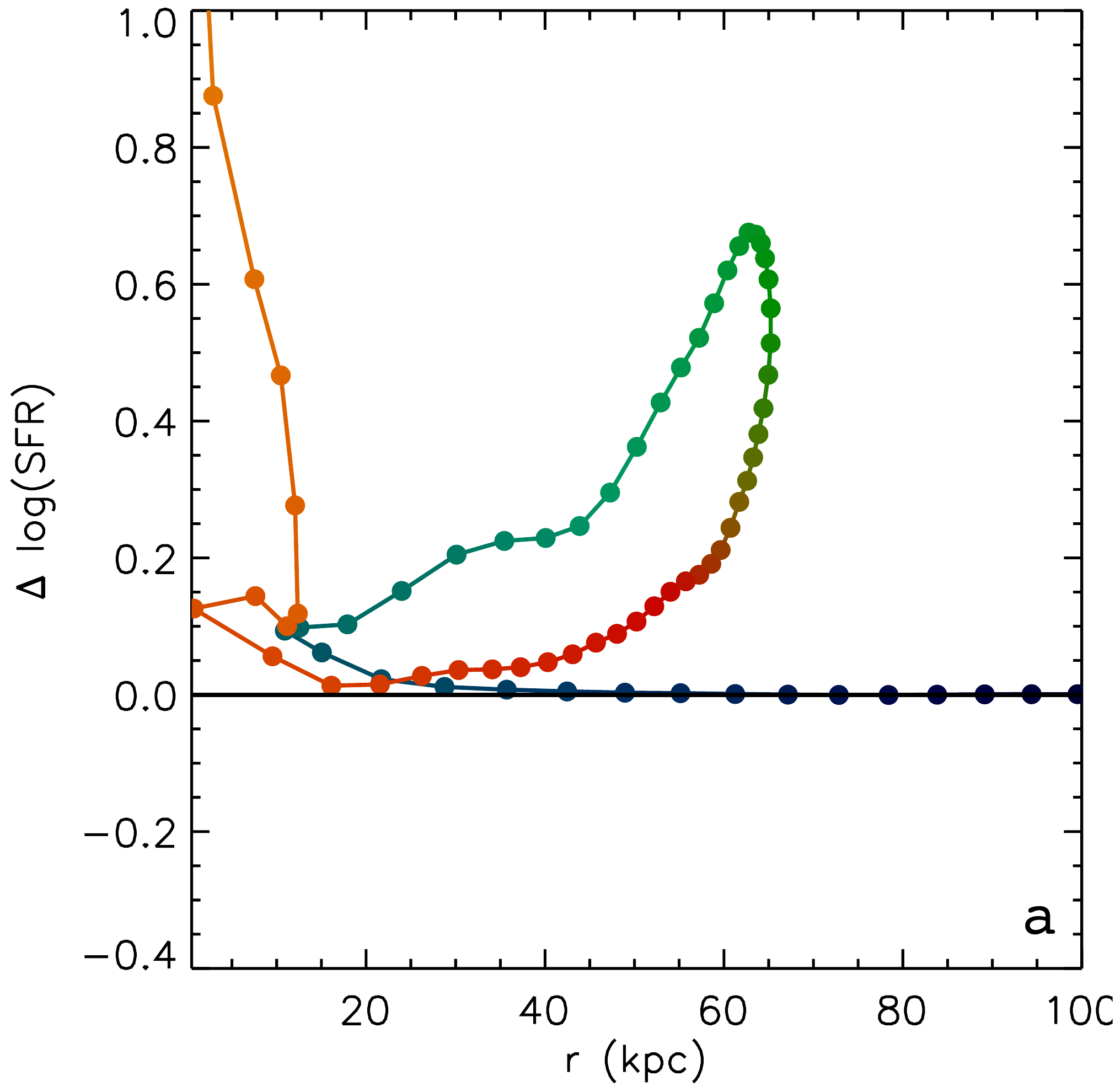}
  \end{minipage}
  \begin{minipage}[c]{8 cm}
\includegraphics[width=220px]{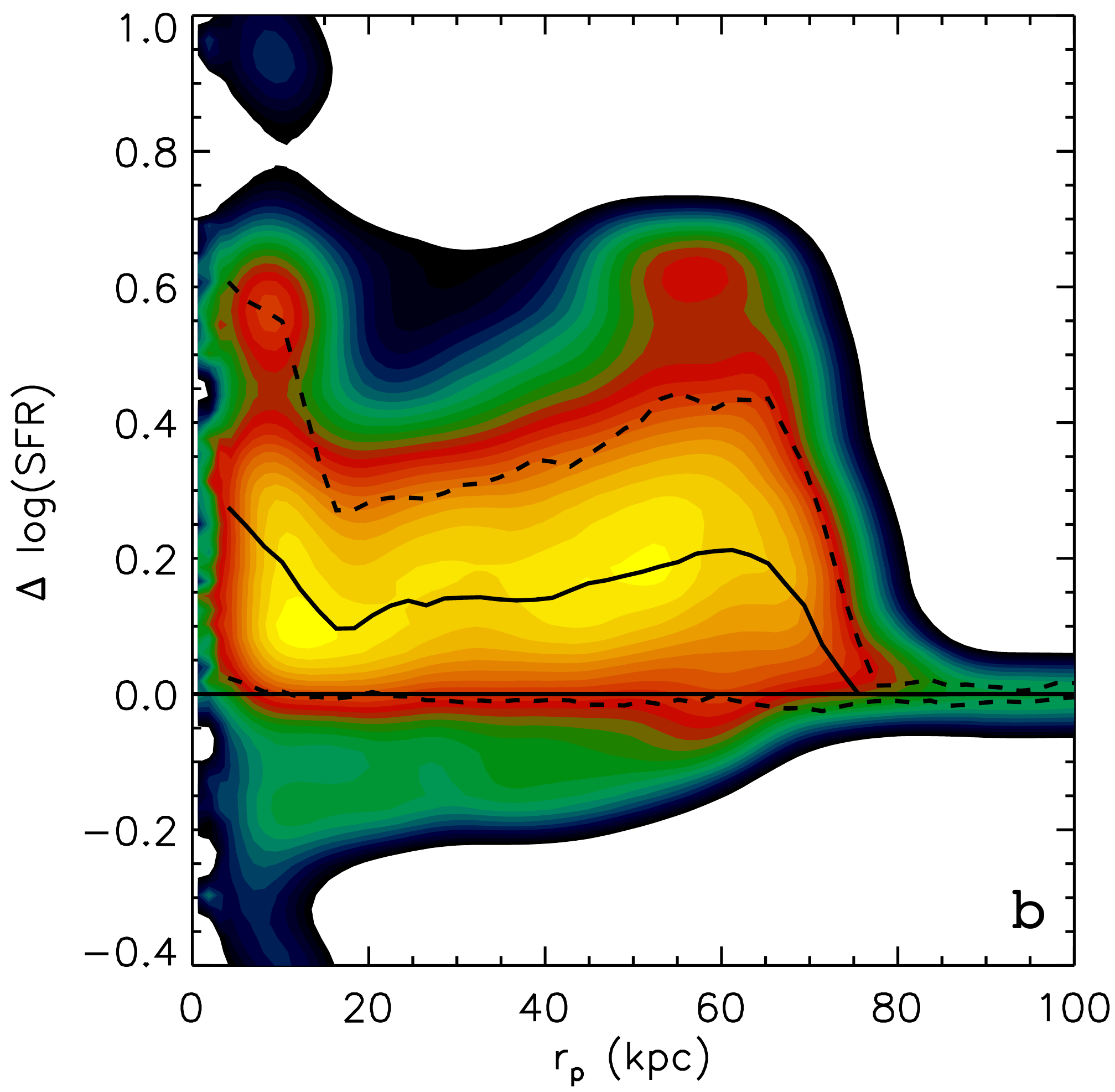}
  \end{minipage}
\caption{\label{fig:sfr_sims} Same as Figure \ref{fig:met_sims}, but for SFR.  The contours in Panel b outline the same general form as is seen observationally, with a wide separation plateau, and higher SFR enhancements at small separations.}
\end{figure*}

In order to more directly compare the results of the simulation to our data, we must fold the \deloh~and \delsfr~tracks into a plot as a function of separation, rather than as a function of time elapsed.  Figures \ref{fig:met_sims}a and \ref{fig:sfr_sims}a show metallicity and SFR tracks respectively for one galaxy's response to a merger, as a function of true physical separation.  The metallicity and SFR are measured at fixed time steps (10 Myrs) throughout the merger, and points are colour-coded as a function of time (from dark blue at early times to yellow at late times).  Figure \ref{fig:met_sims}a shows that the maximal dilution in \deloh~after the first close passage seen in Figure \ref{fig:sim_timeline} occurs at a separation of $\sim$ 60 kpc (in this particular merger), with the final metallicity drop due to coalescence seen at real separations of $<15$ kpc, as the galaxies coalesce.  Similarly, the strongest SFR enhancement after first passage occurs in a relatively narrow peak between $45-65$ kpc, with the coalescent peak appearing at separations $<15$ kpc.  Interestingly, between the two peaks, the SFR appears to drop to nearly its original value; high SFR offsets are generally present at wide separations after a close passage, or at very small separations due to coalescence.  However, the narrowness of the peak in SFR at 60 \kpc~in separation space is a result of the coincidence of the galaxies being at apocentre during the period of time when the SFR enhancement strengthens.  There is no reason to expect the apocentre and the SFR burst to coincide; this is a coincidence of our merger orbital parameters.  If the galaxy takes longer to reach apocentre, the SFR enhancement would begin prior to apocentre, and this SFR peak would have a broader distribution in separation space.

Physical separations are not an observable quantity, so in order to place the results of the simulations on more comparable footing with the data, we introduce the full suite of 16 mergers, and convert the physical separation tracks into projected separation space.  The conversion takes each point from the simulations and multiplies by a random viewing angle in 3-dimensional space, defined as $| \mathrm{cos} (\phi )|$, where $\phi=sin^{-1}(R)$ and $R$ is a different random value between 0 and 1 for each point in the simulations. The cosine of a random angle accounts for the 2 dimensional spin of the viewing angle, and to account for the distribution of angles over the surface of a sphere, we use the arcsine of a random number between 0 and 1, which should properly account for the 3 dimensional distribution of possible viewing angles. 
We can then plot the simulated \deloh~and \delsfr~as a function of projected separation.  The contours of the scattered points are shown in Figures \ref{fig:met_sims}b \& \ref{fig:sfr_sims}b respectively.  The projected separation contours are time-weighted, as each point in the real separation simulations is measured at fixed time intervals.  In regions of real space where the galaxies spend most of their time, there is a corresponding increase in the number of points at that distance.  

Strikingly, these contours outline the same general form as is seen in the SDSS sample of pair galaxies.  The metallicity contours outline a low level plateau of metallicity dilution out to 70 \kpc, with a sharper drop at very small separations.  The contours do not show a marked change from the individual tracks, other than an expected blurring of minor features due to projection effects and the scatter introduced by including several merger orientations.  The SFR enhancements undergo a more dramatic transformation between real and projected space.  The reasonably tight peak in SFR at wide separations has scattered to smaller separations, resulting in a distinct plateau of enhanced SFR out to 70 \kpc.  Converting to projected separations will scatter galaxies to smaller separations, so the wide separation peak has filled in the previous gap between the high \rp~post-pericentre peak and the peak due to coalescence.  

The cutoff of the offset plateau at $\sim70$ \kpc~is a result of our simplified merger suite.  As we do not explore a large range of potential orbits, varying only the galaxy inclination, none of our mergers reach separations beyond $\sim$70 \kpc~after their first pass.  With a wider suite of simulations, we expect this low level plateau would naturally extend as galaxies reach larger distances after their first pass (Patton et al., in prep).  Observationally, this would correspond to galaxies in either more weakly bound interactions than those we have simulated here, or galaxies with higher initial angular momenta.  The residual peak at 60 \kpc~is also a function of the fixed initial orbital energy and angular momentum in all 16 of our simulated mergers; keeping these two parameters the same results in fixed orbits.  With a wider range of orbits, this feature should weaken into a flatter plateau, as galaxies with smaller and wider apocentres blend together.  
Since the strength of the tidal interaction is expected to directly impact the strength of the induced SFR \citep[e.g.,][]{diMatteo2007, Cox2008}, we also expect that with a broader range of tidal interactions, we should see a corresponding increase in the range of SFR enhancements induced in the simulations.

The qualitative similarity between the contour plots resulting from this simple suite of 16 major mergers and the observational results makes the framework of these simulations an appealing one for the interpretation of the observed trends.  In this scenario, the innermost peak in offset values should be due almost entirely to galaxies approaching coalescence, while the wide separation plateau is due to galaxies which have gone through a close passage, and are only showing the SFR and metallicity response at wider separations.  These galaxies could ultimately merge, or be part of a population of fly-by encounters, which are also expected to show SFR enhancements \citep[e.g., ][]{diMatteo2007, Montuori2010}.

It is necessary to caution that these simulations are not intended to function as a direct quantitative comparison to the data, but simply as a theoretical framework to help interpret the form of the signal observed in the SDSS data.   With a basic set of simulations, we have not reproduced (or indeed, attempted to generate) a representative sample of the range of mergers that exist within the SDSS sample.  The galaxies in the simulations are major mergers only, with a single progenitor galaxy, on a constant merging orbit.  We can state only that the simulations do not lie in a region of parameter space devoid of points in the SDSS data.  With these concerns in mind, we do not overplot the SDSS data on the simulation contours, nor do we encourage a direct comparison of the magnitude of these effects.   There are a large number of parameters which could change the magnitude of the signal seen in our suite of simulations (e.g., a varying fraction of interloping or pre-pericentre galaxies, the influence of minor mergers, gas fractions, or initial orbits), many of which have been the subject of other simulation studies  \citep[e.g.,][]{diMatteo2007, Cox2008}.  However, the comparison of the general form of the trends with \rp~seen in the \delsfr~and \deloh~figures is robust.  The fact that the trends seen in the SDSS sample is reproducible using a straightforward set of simulations is encouraging.  This match in form indicates that the simulation does not need to be tuned to the particulars of our sample to observe the same general trends.  

\section{Discussion}	
\label{sec:discussion}
We find that in a strictly selected sample of galaxy pairs from the SDSS DR7, the star formation rates are typically enhanced by at least 30\% out to separations of 80 \kpc.  The metallicities are suppressed by --0.02 dex ($\sim5$\%) within 60 \kpc.  We visually classify all galaxies in our sample for signs of morphological disturbance to further clean the sample of galaxies which have not yet undergone an interaction or of the remaining fraction of interloping galaxies.  When only the disturbed subsample is taken, the metallicity trend increases in significance over the entire range of \rp~and is offset from the control values out to at least 80 \kpc~by $-0.03$ dex.  Within the inner 30 \kpc, the disturbed sample is offset by --0.04 dex (9\%) in metallicity, and enhanced by 65\% in SFR, relative to the control sample.
Although previous studies have found enhanced SFRs and diluted metallicities in samples of close pairs \citep[e.g.,][]{Ellison2008, Kewley2006a}, this is the first time that the changes in SFR and metallicity have been studied simultaneously as a function of projected separation.  

\subsection{Star formation rate enhancements out to 80 kpc.}
Most previous work using large sample statistics has found that the enhancements seen in SFR are restricted to within 30--40 \kpc~separations \citep[e.g.,][]{Xu1991, Lambas2003, Alonso2006, Ellison2008, Robaina2009}.  
Indeed, in our sample, the strongest signals are present at the smallest separations.  However, we find statistically significant offsets in metallicity and SFR out to the maximum separations of our sample (80 \kpc).

Many previous works have used a mean control value as the point of comparison for their SFR enhancements \citep[e.g.,][]{Lambas2003, Alonso2006, Ellison2008, Lambas2012}, whereas we have used ``offsets'' for individually matched galaxies.  We repeat our analysis using the mean of the control sample as the baseline for our SFR enhancements.  Our results are broadly unchanged; we still see significant SFR enhancement at 80 \kpc.  Evidently, the change in methodology is not entirely responsible for the increase in sensitivity to changes in SFR at wide separations.  
If the methodology is not the dominant improvement,  then the other major set of differences between our work and previous works is in the definition of a control sample. 
While most of the previous works match their pair and control samples in redshift, few of them are matched in stellar mass.  As the SFR of a galaxy is strongly dependent on mass, if the distributions of mass between the pairs and the control galaxies are not well matched, this could well influence the strength of the overall SFR enhancement measured.  For instance, \citet{Lambas2003} and \citet{Darg2010} match to a control sample only in redshift.  Other works have used magnitudes as a proxy for mass \citep{Woods2007, Wong2011, Alonso2006}.  However, \citet{Ellison2008} showed that this method can dilute trends with galaxy properties with a mass dependence, such as SFR.  The conversion from magnitude ratios to mass ratios is not 1:1 (see their Figure 3), so taking magnitude ratios as a direct proxy for mass will introduce a significant fraction of galaxies of different masses than intended.  \citet{Perez2009} compares different methods of selecting a control sample, and finds that matching only on luminosity and redshift introduces significant biases into the result, resulting in artificially redder, more massive pairs, which live in higher density regions of space, relative to their controls. The biases introduced by matching in luminosity and redshift can be reduced by 70\% by instead matching a control sample in mass, redshift, and local density.  

Our pairs sample is matched to a control in total stellar mass, redshift, and in local density.  \citet{Patton2011} and \citet{Ellison2008} also match in mass and redshift.  The extent of the SFR offset trend is remarkably similar to the results found in \citet{Patton2011}.  The \citet{Patton2011} study used a superset of our sample, as it did not require the gas phase metallicity calculations, and uses a comparable offset methodology to ours; direct comparisons between the results are robust.  \citet{Patton2011} find that the central ($g-r$) colours of blue cloud galaxies are consistently bluer than the control values out to 80 \kpc~(see their Figure 15).  \citet{Ellison2008} sees a hint of this trend in major mergers, but the errors are large enough that the trend is not systematically significant to the widest separations, and the full sample only shows significant enhancement at \rp$<30$ \kpc.  Our sample is moderately larger than that of the \citet{Ellison2008} work, at 1899 galaxies versus their 1719, and we have a slightly more stringent \delv~requirement of 300 km/s instead of 500 km/s, both of which would help in reducing error bars.  
Furthermore, \citet{Ellison2008} use total star formation rates, rather than the fibre values used in this work.  
 \citet{Patton2011} determined that the strong offsets in colour are primarily a nuclear effect, as the total ($g-r$) colours of the galaxy showed smaller offsets across all \rp~by a factor of $\sim 4$, and lacked the small separation peak to bluer colours.
We use the aperture corrected SFRs provided in \citet{Brinchmann2004}\footnote{\citet{Brinchmann2004} use model fits to the photometry outside the fibre to make the correction from fibre to total values.  A complete discussion of potential biases is presented in that work.} to test whether we see an analogous weakening of the magnitude of the SFR offsets.  Consistent with the \citet{Patton2011} work, we find that the galaxy's total SFR shows a weaker trend than the fibre values, and lacks the central spike to higher SFR.  However, SFR offsets for the total values are still visible at 80 \kpc, so while using the fibre values increases the magnitude of our SFR enhancements, the use of the fibre values alone is not enough to account for the new sensitivity.  A similar comparison is impossible for the metallicities, as there is no correction from fibre metallicities to total metallicities.  The nuclear concentration of the SFR enhancement indicates that it is much more likely to detect strong SFR offsets when looking at the fibre values rather than the total values, as in \citet{Ellison2008}.
We therefore propose that one of the main factors in increasing our sensitivity to small effects in SFR and metallicity is due to our tightly mass-matched control sample.  Further aiding our sensitivity is the use of fibre SFRs rather than the aperture corrected values.

\subsection{Mass Ratios}
As suggested by many previous studies, the form of the trend with projected separation is primarily driven by the contribution of major mergers \citep[e.g.,][]{Woods2006, Woods2007, Ellison2008}.  
 Investigating the more massive and less massive companions in minor mergers (mass ratios more extreme than the 3:1 major merger criterion), we find that both the less massive and more massive companions show SFR enhancement at all \rp.  The more massive galaxies distinguish themselves from the less massive galaxies only in the smallest separations (see Figure \ref{fig:sfr_mratio}).  Major mergers have slightly higher median offsets, but the median value is very comparable to that of the minor merger enhancements.  
 \citet{Lambas2012} find a similar effect; minor mergers show enhanced SFRs at all masses, relative to the control.  Their sample of major mergers is found to have SFRs enhanced at a slightly stronger level than the minor mergers, but in large part the offsets of the two samples are consistent within error bars.  
 We find that minor mergers are just as effective as major mergers at inducing SFR enhancements of less than a factor of 3 over the control.  
  However, the rarest, most extreme starbursts occur almost entirely within major mergers.  Since these extreme starbursts are so rare, our overall offset distributions are consistent between major mergers and minor mergers.

Previous studies of SFR enhancement as a function of mass ratio have drawn conflicting interpretations from their data (e.g., \citealt{Lambas2003} determined that the more massive galaxy in a minor merger is more strongly affected, while \citealt{Woods2007} found the inverse), a closer inspection of the data indicates that these two previous observational results and that of our current work are not inconsistent.
\citet{Woods2007} found no evidence for a correlation between small \rp~and high SFR for the higher mass companion in a minor merger, whereas there was some evidence of correlation for the less massive companion, and conclude that the less massive companion is more strongly affected by a merger than the high mass companion.  
However, since the SFRs of the galaxies in pairs are not directly compared to a control sample, it is impossible to judge whether or not these galaxies are also systematically enhanced over the control value.  For example, our sample of less massive galaxies in a minor merger shows systematic SFR enhancement, but no correlation with the \rp~of the galaxy pair, so on this point our data do not necessarily conflict.

\citet{Woods2007} also found that when comparing the distributions of specific star formation rates (SSFRs), the distributions of both the more massive and less massive galaxies in minor pairs were statistically consistent with the field sample.  \citet{Lambas2003} also find that galaxies in a minor merger show no significant SFR enhancement at any \rp, compared with the average control SFR value.
In contrast, we find that the distribution of star formation rates at a given mass are statistically different for galaxies in minor pairs and our control sample; KS tests indicate that the less massive galaxies are unlikely to be drawn from the same parent distribution as the control at $\sim6\sigma$, and at $>5\sigma$ for the more massive galaxies in a minor merger.  We note that both \citet{Lambas2003} and \citet{Woods2007} are subject to the same issues with matching a control sample in magnitudes instead of using stellar masses described in \citet{Ellison2008}, which will weaken the sensitivity of their measurements.
  
While our results are not in conflict with the observational results, our results do not appear to align with the expectations from simulations.  There are relatively few simulations which have investigated the SFR enhancements of galaxies in unequal mass mergers.  \citet{Cox2008} find that high mass galaxies in unequal mass mergers are less likely to be tidally perturbed and drive strong SFR enhancements when interacting with a low mass companion, whereas the low mass companion will be strongly perturbed by a massive companion.  Both \citet{Cox2008} and \citet{Mihos1994} find that minor mergers can drive some gas inflows in the massive companion, but the inflowing gas is not necessarily converted into stars.  \citet{Mihos1994} find that the massive companions only show enhanced SFR at coalescence, which does not help explain our trends at wide separations.  \citet{Cox2008} is meant as an improvement upon the \citet{Mihos1994} work, and finds that the massive galaxy is unlikely to undergo a starburst except in very specific cases.
If this theoretical model were borne out, we might then expect to see significantly stronger \delsfr~in the less massive companions, particularly at small separations, where the galaxies are likely to be in the final stages of a merger.  In contrast, we find that the more and less massive galaxies show similar levels of SFR enhancement over most of the range in \rp~probed by our sample, with the \textit{more} massive companions displaying higher SFR enhancement at the smallest separations.  

\subsection{Magnitude of the SFR enhancement}
We can next compare the magnitude of our SFR enhancements to those found in previous works. 
We use a series of statistical tests beyond the median values to determine which magnitude of offsets are preferentially found in the pairs sample instead of the control.  We find that close pairs (\rp $<30$ \kpc) preferentially contain the strongest SFR enhancements (up to a factor of 10 over the control), and that galaxies at wider separations show a statistical excess of more moderate enhancements (up to a factor of 7 over the control).   These tests indicate that while the median values are kept at a much lower \delsfr~due to the relative scarcity of extreme offsets, galaxies in pairs preferentially produce strong SFR enhancements.  However, since this is a test unique to this work, the most straightforward comparison to previous studies is a comparison of median values.

The median SFR enhancement in our results is 40\% over the control; this roughly corresponds to the value of the wide separation plateau.  The enhancements at smaller \rp~are between 60\% and 80\% enhancement over the control values.   
The magnitude of the SFR enhancement we find here is broadly consistent with that found in previous studies.  \citet{Robaina2009} find enhancements of order 80\% over the control values for galaxies with \rp$<40$ \kpc, within mass ratios of 1:4, and between redshifts of $0.4 \leq z \leq 0.8$, consistent with the enhancements seen in our inner peak.
\citet{Lin2007} find that galaxies up to $z\sim1$ and \rp$<50$ \kpc~are enhanced by a factor of $1.9 \pm0.4$, relative to a control sample.  This level of offset is also consistent with the SFR enhancements in our sample. 
 \citet{Wong2011} find that the average enhancement is generally 15-20\% for galaxies within 50 \kpc, and 25--30\% for galaxies within 30 \kpc ~for a sample of galaxies $0.25\leq z \leq 0.75$.   These values are slightly lower than even our wide separation plateau enhancements.  However, \citet{Wong2011} use a very generous \delv~cut of \delv$\leq 3000$ \kms, which will introduce a non-negligible fraction of interloping galaxies into their sample, their results are a lower limit.  
 Some differences in the observed SFR enhancement between works may also be introduced based on differences in the method used to obtain the SFR values, as different methods are well known to give different results \citep[e.g.,][and references therein]{Kennicutt2012}.  For instance, \citet{Wong2011} use dust-corrected ultraviolet colours as a tracer of SFR, while \citet{Lin2007} use infrared luminosities to calculate the SFRs.  Both the effectiveness of the dust corrections and the robustness of the control sample will influence how robust the enhancements are to comparisons between studies.
With these caveats in mind, the relative strength of our results is consistent with those of \citet{Wong2011}. 
 It is intriguing that the average SFR enhancement in galaxy pairs is consistent with our work even at redshifts up to $z \sim 1$.

\subsection{Metallicity offsets}
There has been very little work on the dependence of metallicity on projected separation, and certainly none with the statistics we present here.  \citet{Kewley2006a} have a sample of 86 galaxies in pairs with \rp$<50$ \kpc~and found that metallicities were only significantly diluted at separations closer than 20 \kpc.  Galaxy pairs with \rp~$<20$ \kpc~have an average suppression of $-0.2$ dex relative to non-pairs.  However, \citet{Kewley2006a} uses the luminosity-metallicity relation in lieu of a mass-metallicity relation to calculate their changes in metallicity, and \citet{Ellison2008} showed that $\sim50$\% of this offset was likely due to an increase in luminosity due to enhanced SFRs.  While our median values are not nearly as low as those found in \citet{Kewley2006a}, (between $-0.03$ and $-0.04$ dex within 20 \kpc~for our total and disturbed subsamples respectively), the unbinned points in Figure \ref{fig:all_unbinned} show a significant population of galaxies at offsets of $-0.1$ to $-0.2$ dex, with extreme outliers down to $-0.7$ dex.  
Changes in metallicity may also be sensitive to internal galaxy parameters such as the gas fraction, similar to the dependence for SFR suggested by \citet{diMatteo2007}.  Metallicity dilutions have been shown to be strongest in low density environments, with metallicity enhancements present in cluster environments, where the gas fraction is likely to be low \citep{Ellison2009}.  In support of this idea, \citet{Skillman1996} find that the most gas-deficient galaxies in the Virgo cluster were also the most metal-rich.  Including low gas fraction galaxies in the sample may therefore shift the average \deloh~to values closer to the control sample.

Rather than probing metallicity as a function of separation, most prior work has has focused on the mass--metallicity relation in a sample of closer pairs, selecting only galaxies within 25--30 \kpc~\citep{Ellison2008, Michel-Dansac2008} where the effects of the merger are expected to be the most visible.  In this context, it has been demonstrated that galaxies in pairs are systematically lower in metallicity at fixed mass, relative to a non-pair control, and that the magnitude of this effect is approximately $-$0.03--0.05 dex \citep{Ellison2008, Michel-Dansac2008}.  
If we limit our total sample to the galaxies with \rp$<30$ \kpc, we obtain a median \deloh~of $-0.03$ dex; for the disturbed subsample, the median offset increases to $-0.04$ dex.  \citet{Michel-Dansac2008} calculates metallicity offsets in a similar way to this work, using the mean metallicity of a set of control galaxies of similar mass as the zero point.  However, their metallicity offsets are only visible in the strongly merging sample, and not the tidally disturbed sample.  
Our division by morphological disturbances does not distinguish between these two types, so the higher magnitude metallicity offset seen in \citet{Michel-Dansac2008} ($-0.05$ dex vs. $-0.04$ dex) is perhaps unsurprising.  
Therefore, while the form of our trend with \rp~is novel, both the sign and magnitude of the offsets are consistent with previous studies of metallicities in galaxy pairs.

\subsection{Simulations}
That the trends we observe in the SDSS data are broadly reproduced by the simple suite of simulations presented here is evidence that the merger tracks offered by the models may be an accurate framework within which to interpret our results.  We have made no effort to tune the simulations to our data set, using the same galaxy model for all interactions.  As mentioned in Section \ref{sec:sims}, we are not reproducing a representative sample of mergers, or fully exploring the parameter space of the simulations.   For instance, the location of the large separation peak seen in Figure \ref{fig:sfr_sims}a is partially a function of the initial angular momentum given to the galaxy pairs.  Further, the simulations in our simple suite do not include any galaxies which progress further than $\sim$65 \kpc~away from their companion after the first passage; as our plateau continues until at least 80 \kpc, we expect there to be some fraction of galaxies in our sample which are involved in more weakly bound (or unbound) interactions, which can carry the galaxies to much wider separations (Patton et al, in prep).  With these caveats in mind, the fact that we do reproduce a similar trend to the data, but using a minimum of additional assumptions, is very reassuring.  This indicates that the models do not need to precisely reproduce the exact mergers in the data to provide us with a theoretical model of what the galaxies, on a statistical level, are doing as they progress through a merger.  Using these models as an interpretive tool, the increase in the magnitude of offsets in the smallest \rp ~bins is due to the sharp spike seen due to galaxies near the end of their merger at those separations, when large amounts of metal poor gas are dumped into the nuclear region of a galaxy, and a large SFR burst is triggered.  This dramatic peak due to coalescence at small separations has been seen before in both theoretical \citep[e.g.,][]{diMatteo2007, Montuori2010, Torrey2012} and observational works \citep[e.g.,][]{Larson1978, Donzelli1997, Barton2000, Lambas2003, Alonso2004, Alonso2006, Woods2007, Ellison2008, Ellison2010, Darg2010, Xu2010}.
By contrast, the wide separation offsets are due to a combination of galaxies making their way out after first passage, or possibly back in for a second or final passage, depending on the orbits of the interacting galaxies.  The influence of projected pairs and galaxies that have not yet interacted would result in an overall lowering of the magnitude of the offsets at all \rp.

\section{Conclusions}
\label{sec:conclusions}
We have used a sample of 1899 close pairs from the Sloan Digital Sky Survey's Data Release 7 to study SFR and metallicity offsets as a function of projected separation.  We use a simple suite of simulations to interpret our results.
The main conclusions of this work are as follows:
\begin{enumerate}
\item{Galaxies undergoing an interaction show significant metallicity depressions and SFR enhancements, relative to a control sample that is tightly matched in stellar mass, redshift, and local density.  Galaxy pairs show enhanced SFRs out to projected separations of 80 \kpc, the widest separations in our sample, while significant metallicity dilution is observed out to $\sim$60 \kpc.  Within 30 \kpc, SFRs are enhanced, on average, by $\sim$0.21 dex, or by 60\%, over the control.  At separations wider than 30~\kpc, the SFRs are enhanced by $\sim$0.1 dex (25\% above the control). Metallicities are found to be diluted by --0.02 dex ($\sim$5\% lower than the control) within 60~\kpc.}
\item{When only the morphologically disturbed subsample of galaxies is taken, the metallicity offsets are significantly offset from the control out to 80 \kpc, and the median offset within 30 \kpc~drops to --0.04 dex ($\sim$9\% lower than the control).  Galaxies are visually classified as either visibly disturbed or not visibly disturbed to attempt to reduce the fraction of galaxies which have not yet interacted.}
\item{The form of the \delsfr~and \deloh~offsets as a function of \rp~are primarily driven by galaxies in major mergers.  Both the more massive and less massive companions in a minor merger show similar SFR enhancements to each other and to the major mergers.  The less massive companions show a minor enhancement at all \rp, but do not show the same increased SFR enhancement at small \rp~as is visible in the major mergers or more massive companions.  Furthermore, we find that the most extreme offsets are preferentially found in major mergers, but that minor mergers are equally effective at inducing moderate starbursts in both the less and more massive companions.}
\item{We find that galaxies with strong starbursts, e.g., 10 times as strong as the control, are relatively rare, occurring in only 3\% of galaxy pairs, but are 3 times as likely to be found in the close pairs sample (separations $<30$ \kpc) than in the wide pairs or control sample.  Extremely diluted metallicities (e.g., 1.8 times as metal poor as the control) are rare (3\%), but are twice as likely to be found in the close pairs sample than in the control.  There is evidence for synchronous starbursts as a result of a galaxy pair's interaction.  Galaxies in pairs are at least 17\% more likely to show significantly enhanced SFR (relative to a scrambled pairs sample) if their companion galaxy is also enhanced.}
\item{We use a simple suite of major merger simulations \citep{Torrey2012} to construct a new interpretation of how interactions affect the SFRs and metallicities of a galaxy, as a function of projected separation.  We calculate the \delsfr~and \deloh~offsets from the simulations for a qualitative comparison with the observational trends.  The models are able to reproduce the small separation peak in offset value and wider separation offset plateau seen in the SDSS data. The simulations offer the interpretation that the wide separation plateau is caused by projection effects blurring post-pericentre starbursts at wide separations to smaller separations.  The large offset peak at small separations, by contrast, is primarily due to the extreme starburst induced at the end of a merger, near coalescence, and is less subject to projection effects.}

\end{enumerate}

Galaxy mergers clearly induce large scale gas inflows in galaxies.  However, our observations, combined with the theoretical framework of hydrodynamical simulations, indicates that these inflows occur on longer timescales than has been previously assumed.  With metal poor gas reaching the central regions of a galaxy at a slower rate, the nuclear gas-phase metallicities will take longer to dilute significantly, and the gas reservoir necessary to fuel a significant starburst will take longer to accumulate.  By the time these signatures of gas flow in a galaxy are measurable after the first encounter, the two galaxies have progressed out to wide separations.  As the galaxies reach the end of their merger, the tidal forces inducing these gas inflows become significantly stronger, and the gas phase metallicities and star formation rates change both more rapidly and more dramatically.
Future work will quantify the extent of separations over which it is possible to see the effects of an interaction, and attempt to determine what parameters govern the highest SFR galaxies.

\section*{Acknowledgments}
We thank the anonymous referee for a constructive report.
We are grateful to the MPA/JHU group for access to their data products and catalogues (maintained by Jarle Brinchmann at  \url{http://www.mpa-garching.mpg.de/SDSS/}).   SLE and DRP acknowledge the receipt of an NSERC Discovery grant which funded this research.

Funding for the SDSS and SDSS-II has been provided by the Alfred P. Sloan Foundation, the Participating Institutions, the National Science Foundation, the U.S. Department of Energy, the National Aeronautics and Space Administration, the Japanese Monbukagakusho, the Max Planck Society, and the Higher Education Funding Council for England. The SDSS Web Site is \url{http://www.sdss.org/}.

The SDSS is managed by the Astrophysical Research Consortium for the Participating Institutions. The Participating Institutions are the American Museum of Natural History, Astrophysical Institute Potsdam, University of Basel, University of Cambridge, Case Western Reserve University, University of Chicago, Drexel University, Fermilab, the Institute for Advanced Study, the Japan Participation Group, Johns Hopkins University, the Joint Institute for Nuclear Astrophysics, the Kavli Institute for Particle Astrophysics and Cosmology, the Korean Scientist Group, the Chinese Academy of Sciences (LAMOST), Los Alamos National Laboratory, the Max-Planck-Institute for Astronomy (MPIA), the Max-Planck-Institute for Astrophysics (MPA), New Mexico State University, Ohio State University, University of Pittsburgh, University of Portsmouth, Princeton University, the United States Naval Observatory, and the University of Washington.

\bibliographystyle{apj}
\bibliography{masterfile_bibdesk}
\begin{appendix}

\end{appendix}

\end{document}